\documentclass[%
 reprint,
showkeys,
 amsmath,amssymb,
 aps,
]{revtex4-1}

\usepackage{graphicx}
\usepackage{dcolumn}
\usepackage{bm}

\renewcommand{\vec}[1]{\mbox{\boldmath $#1$}}
\renewcommand{\tensor}[1]{\mbox{\boldmath $#1$}}

\newcommand{\upd}{\mathrm{d}}

\begin{document}


\title{
Improving mixing characteristics with a pitched tip 
in kneading elements in twin-screw extrusion
}%

\author{Yasuya Nakayama\(^{1}\)}
\email{nakayama@chem-eng.kyushu-u.ac.jp}
\author{Hiroki Takemitsu\(^{1}\)}
\author{Toshihisa Kajiwara\(^{1}\)}
\author{Koichi Kimura\(^{2}\)}
\author{Takahide Takeuchi\(^{2}\)}
\author{Hideki Tomiyama\(^{2}\)}
\affiliation{%
\(^{1}\)Department of Chemical Engineering,
Kyushu University,
Nishi-ku,
Fukuoka 819-0395,
Japan
}%
\affiliation{%
\(^{2}\)Hiroshima Plant, The Japan Steel Works Ltd. 1-6-1 Funakoshi-minami, Hiroshima 736-8602, Japan
}


\date{\today}

\begin{abstract}
In twin-screw extrusion, the geometry of a mixing element mainly 
determines the basic flow pattern, which eventually affects the 
mixing ability as well as the dispersive ability of the mixing 
element.
The effects of geometrical modification, with both forward and 
backward pitched tips, of a conventional forwarding kneading 
disks element (FKD) in the pitched-tip kneading disks element on 
the flow pattern and mixing characteristics are discussed.
Numerical simulations of fully-filled, non-isothermal polymer melt 
flow in the melt-mixing zone were performed, and the flow pattern 
structure and the tracer trajectories were investigated.
The pitched tips largely affects the inter-disc fluid transport, 
which is mainly responsible for mixing. 
These changes in the local flow pattern are analyzed 
by the distribution of the strain-rate state.
The distribution of the finite-time Lyapunov exponent
reveals a large inhomogeneity of the mixing in FKD is suppressed 
both by the forward and backward tips.
By the forward tips on FKD, the mixing ability is relatively 
suppressed compared to FKD, whereas for the backward tips on FKD, 
the mixing ability is enhanced while maintaining the same level 
of dispersion efficiency as FKD. 
From these results, the pitched tips on the conventional KD turns 
out to be effective at reducing the inhomogeneity of the mixing 
and tuning the overall mixing performance.
%
\end{abstract}

\keywords{Polymer processing,
Mixing,
Twin-screw extrusion,
Finite-time Lyapunov exponent
}
\maketitle


%
\section{Introduction}
The mixing of highly viscous materials, as well as low-viscous liquids, 
is one of the most important processes in chemical engineering.
Twin-screw devices such as twin-rotor mixers and twin-screw 
extruders are widely applied in the mixing processes of highly 
viscous materials, including polymer processing, rubber 
compounding, and food 
processing~\cite{Tadmor2006Principles,Kohlgruber2007CoRotating,White2010Twin,2011Food,Rauwendaal2014Polymer}.
For twin-screw extrusion, various types of mixing 
elements have been developed for different mixing qualities as 
well as different processabilities of the materials, since the 
geometry of the mixing elements mainly determines the 
characteristics of the mixing 
process~\cite{Kohlgruber2007CoRotating}.
In the development of a mixing element, one fundamental issue is 
a systematic understanding of the relation between the flow pattern 
driven by the mixing element and the mixing characteristics.

To understand the mixing abilities of different elements, 
experimental and numerical analyses of the flow in twin-screw 
extrusion have been 
performed
~\cite{Christiano1993J,Lawal1995Mechanisms,Lawal1995Simulation,Cheng1997Study,Carneiro1999Flow,Shearer1999Analysis,Alsteens2004Parametric,Jaffer2000Experimental,Bravo2000Numerical,Shearer2000Effects,Shearer2001Distributive,Shearer2001Relationship,Nakayama2011Meltmixing,Kubik2012Method,Hirata2014Effectiveness,Zhang2009Numerical,SarhangiFard2013Simulation,Yamada2015Dispersive,Nakayama2016Strain,Nakayama2016Effects}. 
A most common mixing element is 
the kneading block or kneading discs element~(KD), which is 
composed of the several oval discs
combined with a certain stagger 
angle~\cite{Kohlgruber2007CoRotating}.
Mixing with kneading blocks in twin-screw extruders
has been experimentally assessed
by analyzing the distribution of tracers or interfaces.
Christiano analyzed the distribution of carbon black in the extrudates
with the intensity of segregation~\cite{Christiano1993J}.
Shearer and 
Tzoganakis~\cite{Shearer1999Analysis,Shearer2000Effects,Shearer2001Distributive,Shearer2001Relationship} 
used a reactive tracer to quantify the interface generation by 
distributive mixing.
They reported that forward KDs have good distributive mixing 
in comparison with reverse and neutral KDs~\cite{Shearer1999Analysis,Shearer2000Effects,Shearer2001Relationship}.
They also reported that the correlation between the mean
residence time and the mixing ability holds for the reverse and 
neutral KD while the forward KD have a different dependence on 
the residence time~\cite{Shearer2000Effects}.
Numerical simulations have been applied to analyze the flow and 
mixing in the kneading block zone.
The important effect of the boundary conditions on the flow in the 
kneading block zone was suggested~\cite{Jaffer2000Experimental}.
Bravo, Hrymak, and Wright reported a comparison between the 
results from the numerical simulation and experimental 
one~\cite{McCulloughSPE}, showing a good agreement of the 
pressure and velocity fields in most of the KD 
zone~\cite{Bravo2000Numerical}.
By modifying the angle of the tips in each disc of KD, a 
refined geometry called ``pitched-tip kneading disks 
element'' has been developed.
In previous 
works~\cite{Nakayama2011Meltmixing,Nakayama2016Effects}, the 
different mixing characteristics of ptKDs have been discussed by 
using the numerical simulations of the flow in the kneading block
zone.
It was reported that, under an operational condition, among 
different ptKDs, the combination of opposite stagger and tip 
angle directions make the residence time distribution broader 
compared to ones with the same direction of the stagger and tip angles.
In addition, 
the mean stress during residence is mainly determined by the 
disc-stagger angle.
Although the prior works showed the different mixing 
characteristics of different ptKDs, suggesting a greater design space of 
the kneading blocks, the advantages of ptKDs to the 
conventional KD are unknown.
The essential effects of the geometrical modification by the 
pitched tips on the flow pattern and resulting advantages over 
the conventional KD still need to be clarified.

In this article, for a conventional forward KD~(FKD) in twin-screw 
extrusion, we investigate how the geometrical modification with 
pitched tips alters the flow pattern and eventual mixing 
quality.
We consider two different tip angles, namely forward and backward 
tip angles, on FKD as well as the conventional one, and discuss 
the differences in the flow patterns and the mixing abilities of 
the different geometries using numerical simulations of the flow 
in the kneading block zone.
In order to discuss the flow pattern structures by different 
kneading blocks, we observe the strain-rate state 
distribution~\cite{Nakayama2016Strain} as well as the velocity 
field.
For quantification of the distributive mixing, we used the 
statistical distribution of the finite-time Lyapunov exponent in 
addition to the residence time distribution.

\section{Geometric structures of pitched-tip kneading discs 
elements}
\label{sec:ptkd}
We discuss three kinds of the kneading blocks~(KD).
One is a conventional type shown in Fig.~\ref{fig:geom}(a) called 
a forward kneading discs element~(FKD)
in which the five discs with non-pitched tips are arranged 
with a stagger angle between each two adjacent discs.
The other geometries used in this study are the two different 
pitched-tip kneading discs~(ptKD) element shown in 
Figs.~\ref{fig:geom}(b) and (c).
In Figs.~\ref{fig:geom}(b) and (c), five discs are arranged 
with a forwarding disc-stagger angle, but the tips of each disc 
are twisted with respect to the screw axes.
Depending on the direction of the tip angle, each disc can 
have a forward or backward pumping ability and thus modify the 
overall pumping ability of the kneading block. 
Corresponding to 
the tip angle directions, we call these pitched tips 
``forward tip''~(Ft) or``backward tip''~(Bt).
Since in Fig.~\ref{fig:geom}(b) the discs with Ft are combined 
with a forwarding disc-stagger angle, we call this geometry Fs-Ft 
ptKD, while in Fig.~\ref{fig:geom}(c) the discs with Bt are 
combined with a forwarding disc-stagger angle, and so we call 
this geometry Fs-Bt ptKD.
Along with this terminology, since FKD uses a non-pitched, neutral 
tip~(Nt), it can also be called Fs-Nt ptKD.
The diameter of the barrel, \(D\), is set to 28\;mm. All the 
discs composing the different kneading blocks have a diameter of  
0.9893\(D\) and a width of 0.286\(D\).
disc-stagger angles and pitched-tip angles are chosen so that the 
both inlet and outlet sections of the kneading blocks coincide as 
shown in Fig.\ref{fig:geom}.
disc-stagger angles for the three kneading blocks were 
respectively set to 45\(^{\circ}\) for FKD, 40.715\(^{\circ}\) 
for Fs-Ft ptKD, and 49.285\(^{\circ}\) for Fs-Bt ptKD.

The pitched tips modify the channel geometry of the conventional 
KD as well as the pumping ability associated with the screw 
rotation.
These effects result in a modification of the flow pattern and 
global mixing.
In this paper, we investigate the relation between the geometry 
and the flow pattern, and discuss how the mixing characteristics 
are modified by this geometrical modification using the different 
pitched tips.

\newcommand{\figone}{
Top view of the three types of kneading blocks discussed in 
this paper. (a) forward kneading discs~(FKD), 
(b) forward-tip discs are combined with forward disc-stagger angle, 
which is called forward stagger and forward tip (Fs-Ft) ptKD, 
and
(c) 
backward-tip discs are combined with forward disc-stagger angle, 
which is called forward stagger and backward tip (Fs-Bt) ptKD. 
}
\begin{figure}[htbp]
\raggedright
 \centering
 \includegraphics[width=1.\hsize]{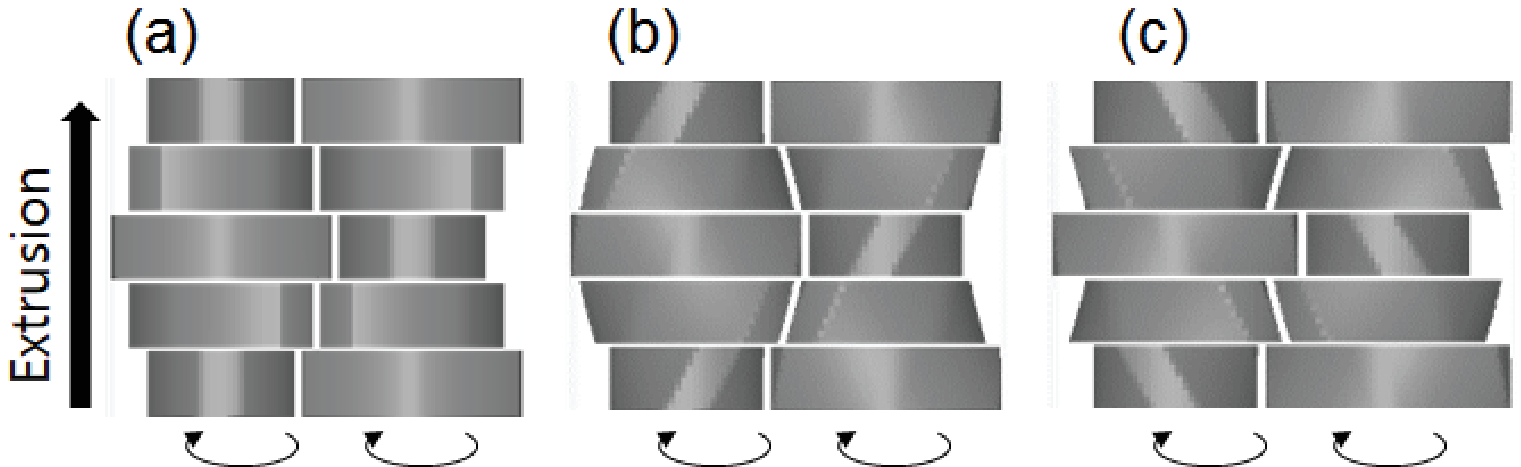}
\caption{
\label{fig:geom}
\texttt{\figone}
}
\end{figure}

\section{Numerical simulation}
\subsection{Screw configuration}
The flow of a polymer melt in the melt-mixing zone of a 
twin-screw extruder has been numerically solved to allow 
understanding the relation between the geometric structure of the 
kneading discs elements, the flow patterns, and the melt-mixing 
characteristics.
The configuration of the screws is shown in 
Fig.~\ref{fig:screw_configuration}.
From the inlet, a forward conveying element, kneading block, and
a backward conveying element are arranged.
The length of the computational domain is \(3.28\,D\).
For the kneading block zone, we have set three types of 
pitched-tip kneading blocks of the \(L/D=1.5\) described in 
the section ``Geometric structures of pitched-tip kneading 
discs elements''.
\newcommand{\figtwo}{Screw configuration of the melt-mixing zone. 
From the inlet (bottom of the figure), forward conveying element, 
kneading block,  and backward conveying element are configured.
The extrusion direction is indicated by the arrow on
the left, while the rotation direction is indicated by
the arrows on the bottom. 
}
\begin{figure}[htbp]
\raggedright
 \centering
 \includegraphics[width=.8\hsize]{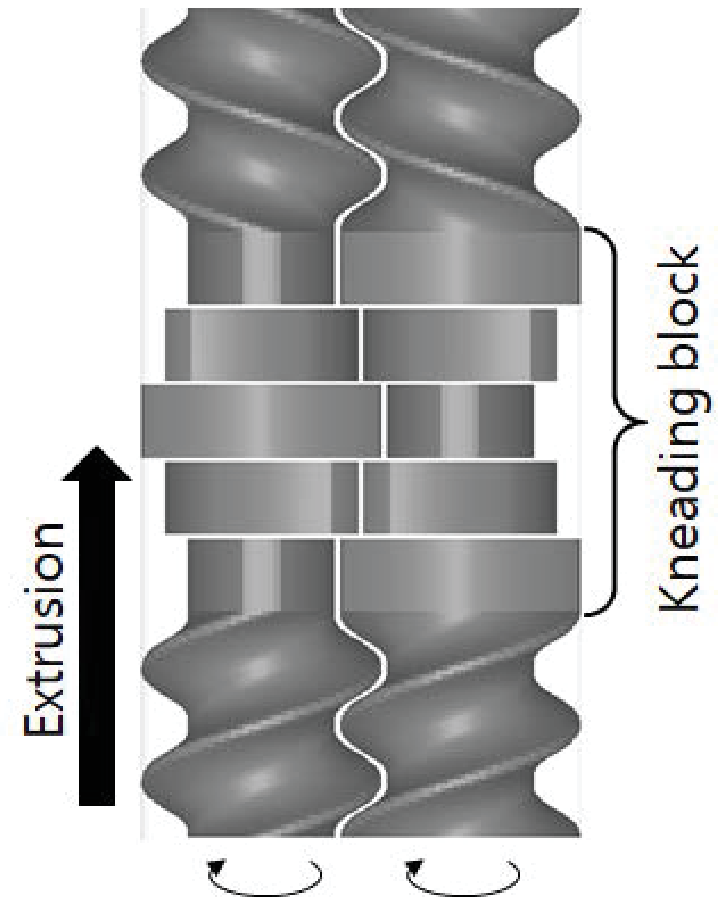}
\caption{
\label{fig:screw_configuration}
\texttt{\figtwo}
}
\end{figure}

\subsection{Working fluid and boundary conditions}
We focus on the situation where the material fully fills the channel.
The Reynolds number is assumed to be much less than unity, so that
inertial effects are neglected.
The flow is assumed to be incompressible, and in a pseudo-steady 
state to the screw rotation, as has often been assumed in polymer 
flow in twin-screw 
extruders~\cite{Ishikawa20003D,Bravo2004Study,Malik20053D,Zhang2009Numerical,Nakayama2011Meltmixing,SarhangiFard2013Simulation,Rathod2013Effect,Hirata2014Effectiveness}.
With these assumptions, the governing equations become
\begin{align}
 \vec{\nabla}\cdot\vec{v} &= 0, 
\\
0 &= -\vec{\nabla}p+ \vec{\nabla}\cdot\tensor{\tau},
\\
\rho c_{p}\vec{v}\cdot\vec{\nabla}T &= k\nabla^{2}T + \tensor{\tau}:\tensor{D},
\end{align}
where \(\vec{v}\) is the velocity, \(p\) is the pressure, 
\(\tensor{\tau}\) is the deviatoric stress, \(\rho\) is the mass 
density, \(c_{p}\) is the specific heat capacity, \(T\) is the 
temperature, \(k\) is the thermal conductivity, and 
\(\tensor{D}=\left(\nabla\vec{v}+\nabla\vec{v}^{\text{T}}\right)/
2\) is the strain-rate tensor in which the superscript T represents 
tensor transpose.

The fluid is assumed to be a viscous shear-thinning fluid that follows
Cross--Arrhenius viscosity~\cite{Cross1965Rheology}, 
\begin{align}
 \tensor{\tau} &= 2\eta\tensor{D},
\\
\label{eq:cross_model}
\eta &= \frac{\eta_{0}}{1+c\left(\eta_{0}\dot{\gamma}\right)^{n}},
\\
\label{eq:arrhenius_model}
\eta_{0} &= a\exp\left(\frac{b}{T}\right),
\\
\label{eq:shear_rate}
\dot{\gamma} &= \sqrt{2\tensor{D}:\tensor{D}},
\end{align}
whose parameters are
obtained by fitting the shear viscosity of a polypropylene melt 
taken
from Ref.~\cite{Ishikawa20003D} , and the values are \(c=1.3575\times
10^{-3}\), \(n=0.66921\), \(a=1.7394\)\;Pa\(\cdot\)s, and \(b=4656.8\)\;K.
The mass density, specific heat capacity, and thermal conductivity are
taken from Ref.~\cite{Ishikawa20003D} as well, and the values are
\(\rho=735\)\;kg/m\(^{3}\), \(c_{p}=2100\)\;J/(kg\(\cdot\)K), and
\(k=0.15\)\;W/(m\(\cdot\)K).  
As operational conditions, the volume flow rate and the screw 
rotation speed are set to 10\;cm\(^{3}\)/s (\(\approx 
26.5\)\;kg/h) and 200\;rpm.  
The no-slip condition on the velocity at the barrel and screw
surfaces is assumed. 
To circumvent the effects of the inlet and outlet boundary 
conditions, we set additional flight-less zones before and after 
the screw region of interest depicted in 
Fig.~\ref{fig:screw_configuration}.
The inlet and outlet boundary conditions are imposed at the ends 
of the flight-less zones.
At the inlet plane, the uniform axial velocity was set to a value 
under the given volume flow rate.
The pressure at the outlet boundary was fixed to be a constant 
value. The temperatures on the barrel surface and at the inlet 
boundary were set to be a constant value of 473.15\;K. The 
natural boundary conditions for the temperature equation in the 
exit boundary plane and the screw surface were assigned.
Physically, in the upstream flight-less zone, the flow develops 
before reaching the forward conveying element.
The length of the flight-less screw zone was chosen to be 
0.071\(D\) so that the flow in the kneading block zone does not 
change. 
To obtain the pseudo-steady solution along the screw motion, the 
steady flow was solved at each three degree of the screw rotation. 
For each angle, the mesh conforming with the screw and barrel 
surfaces is generated. Fig.~\ref{fig:mesh} shows an example of 
the meshing used in the numerical analysis.
\newcommand{\figb}{
Typical meshing conforming the screw and barrel surfaces used in 
the numerical analysis.
The section of the self-wiping screws is generated based on the 
work done by Booy~\cite{Booy1978Geometry}.
}
\begin{figure}[htbp]
\raggedright
 \centering
 \includegraphics[width=1.\hsize]{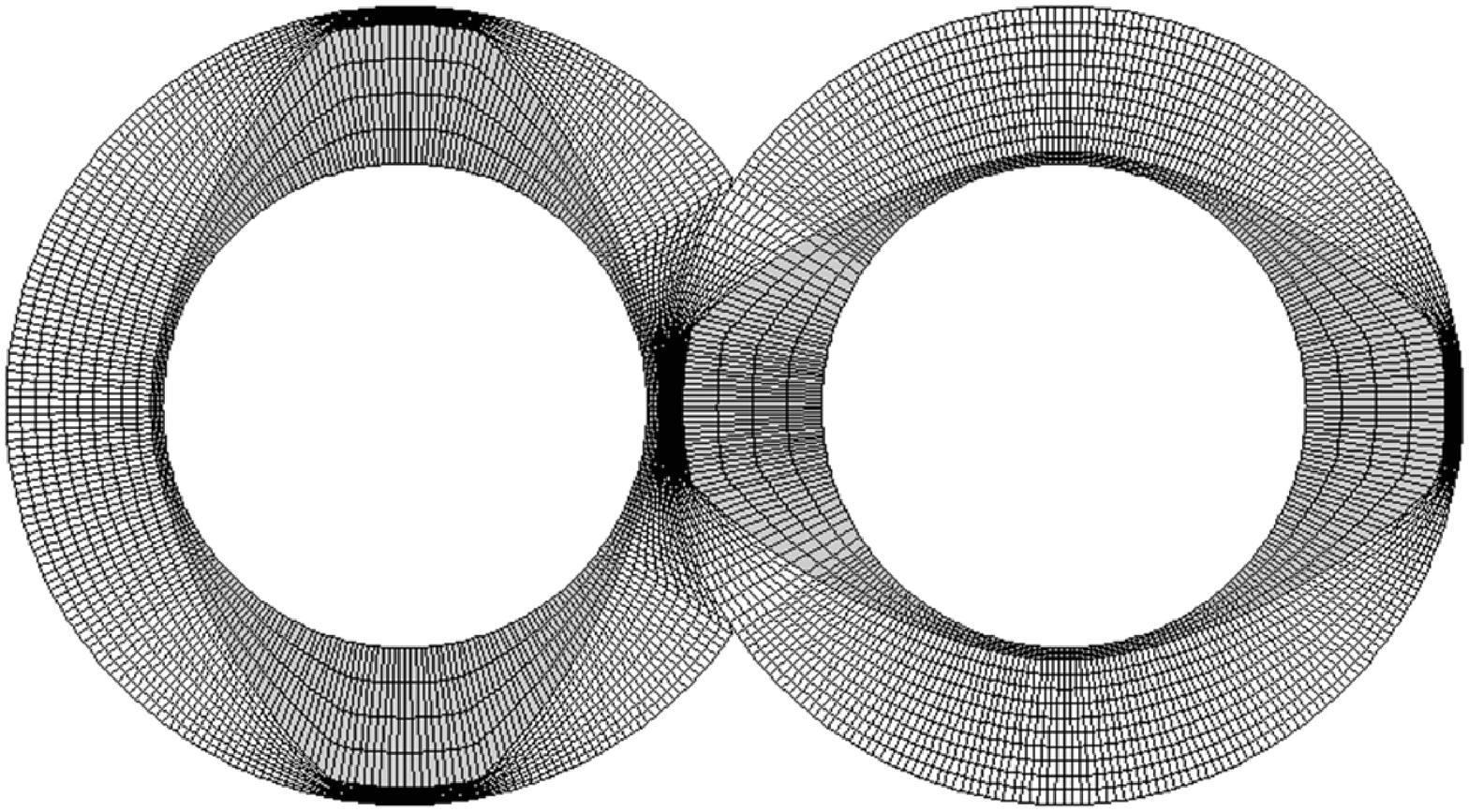}
\caption{
\label{fig:mesh}
\texttt{\figb}
}
\end{figure}

\subsection{Tracer statistics}
The set of equations was discretized by the finite volume
method and solved by SIMPLE method~\cite{Versteeg2007Introduction}
using a commercial software, ``R-FLOW'' (R-flow Co., Ltd., Saitama,
Japan).
The trajectories of the passive tracers were calculated by 
integrating the solved velocity field. 
The set of tracer trajectories allows us to estimate the 
statistical distribution of the flow history, from which we can 
discuss the mixing characteristics in the local region of interest.
This analysis was performed using the codes developed by the 
authors.
The Lagrangian average of a quantity \(f\) over the trajectory of
\(\alpha\)th tracer is defined as
\begin{align}
 \overline{f_{\alpha}}^{t_{\alpha}} &=\frac{1}{t_{\alpha}}\int_{0}^{t_{\alpha}}
\!\!\!\upd
 u\int\!\upd\vec{x}\delta\left(
\vec{x}-\vec{X}_{\alpha}(u)
\right)f(\vec{x},u),
\label{eq:lagrangian_average_of_f}
\end{align}
where \(t_{\alpha}\) and \(\vec{X}_{\alpha}(.)\) are the local 
residence time of the \(\alpha\)th tracer in the kneading 
block zone and the position of the \(\alpha\)th tracer, 
respectively, and \(\delta(.)\) is the Dirac delta distribution.
The statistical distribution of 
\(\overline{f_{\alpha}}^{t_{\alpha}}\) characterizes the 
overall property of the fluid process in the melt-mixing 
zone, while the individual trajectories are connected to the 
local flow.
Initially, about 2600 points were uniformly distributed in a 
certain section in front of the kneading block and were advected 
until they reached the end of the kneading block. 
\section{Results and discussion}

\subsection{Pumping ability}
The geometry of a mixing element defines not only the channel 
geometry, but also the direction and the strength of the pressure 
flow under a flow rate and a screw rotation speed.
The pressure flow combined with the drag flow by the screw 
rotation determines the flow pattern in the mixing zone. 
In ptKDs, the pumping ability is mainly determined by the 
combination of the disc-stagger angle and the tip angle.
In Fig.~\ref{fig:pressure_drop}, the average pressure drop across 
the kneading block
as a function of the screw rotation speed under 
a flow rate of \(Q=10\)\;cm\(^{3}\)/s is drawn for different 
kneading blocks.
In the calculation of the average pressure drop, the 
pressure at the inlet and outlet sections of the kneading block 
is averaged spatially over the sections and temporally over one 
screw rotation.
A positive/negative value of the pressure drop indicates that the 
pressure flow occurs on average in the forward/backward extrusion 
direction.
Figure~\ref{fig:pressure_drop} clearly shows that the pumping 
ability increases in the order of Fs-Bt ptKD, FKD, and 
Fs-Ft ptKD, which reflects the forward and backward drags 
of the Ft and Bt, respectively.
In particular, the Fs-Bt ptKD shows a positive pressure drop, 
indicating a negative pumping ability of Fs-Bt ptKD in 
contrast to the positive pumping abilities of Fs-Ft ptKD 
and FKD.

\newcommand{\figfour}{Average pressure drop in the kneading block 
zone 
over a section and one screw rotation under a flow rate of 10\;cm\(^{3}\)/s as 
a function of screw rotation speed.}
\begin{figure}[htbp]
\raggedright
 \centering
 \includegraphics[width=1.1\hsize]{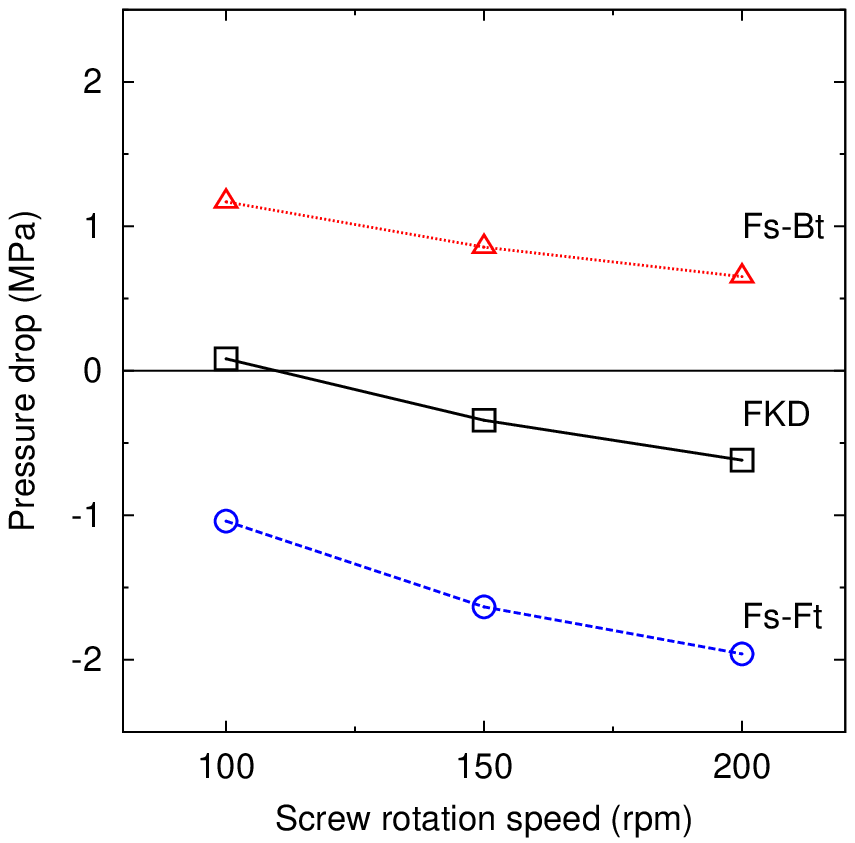}
\caption{
\label{fig:pressure_drop}
\texttt{\figfour}
}
\end{figure}

\subsection{Temperature profile}
Fig.~\ref{fig:axial_temperature} shows the average temperature profile along the 
extrusion direction, where \(\langle ... \rangle_{z}\) represents 
the average spatially over a section specified by an axial 
position of \(z\) and temporally over one screw rotation.
For the three different kneading blocks, the average temperature 
increases with the axial position, which is mainly due to the 
combined effects of the viscous heating and the mixing, but the 
variation of \(\langle T\rangle_{z}\) through the kneading block 
zone 
is rather small. 
Difference in \(\langle T\rangle_{z}\) for different geometries is 
at most 5~K.
From this result, we assume that the temperature does not have a 
significant effect on the mixing ability.
\newcommand{\figa}{
Average temperature over a section 
and one screw rotation as a function of  the axial 
position with a flow rate of 10\;cm\(^{3}\)/s and a screw 
rotation speed of 200\;rpm for FKD (black solid line), Fs-Ft 
ptKD (blue dashed line), and Fs-Bt ptKD (red dotted 
line). 
The vertical dashed lines indicate the locations of the ends of 
the discs in the KDs.
}
\begin{figure}[htbp]
\raggedright
 \centering
 \includegraphics[width=1.1\hsize]{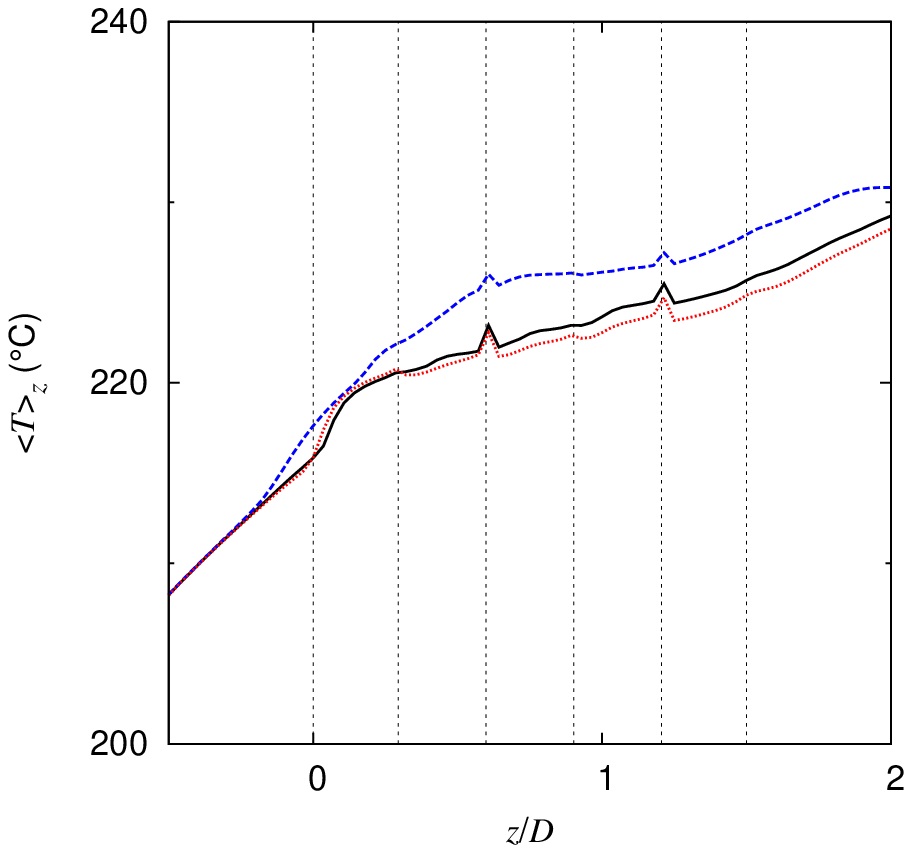}
\caption{
\label{fig:axial_temperature}
\texttt{\figa}
}
\end{figure}

\subsection{Flow pattern}
We discuss the flow patterns generated by different KDs.
Figure~\ref{fig:fkd_flow}(a) shows the velocity field in several 
axial sections in the kneading block zone of FKD, 
Fig.~\ref{fig:fkd_flow}(b) shows the distribution of the 
strain-rate state, \(\beta\) (Eq.~(\ref{eq:beta})), at a mid-section between the barrel and screw surfaces in the 
kneading block zone of FKD.
This section is specified by a mesh number in \(r\)-direction 
shown in Fig.~\ref{fig:mesh} and therefore reflects the 
shape of the screw surfaces.
Fig.~\ref{fig:fkd_flow}(c) shows \(\beta\) distribution in the 
axial cross section indicated in Fig.~\ref{fig:fkd_flow}(b).
The distribution of the strain-rate state reflects the local 
flow pattern and therefore is useful for discussing the relation 
between the geometry and the flow 
pattern~\cite{Nakayama2016Strain}.
A brief description of the strain-rate state is given in 
\ref{sec:stran_state}.

Basically, as the screws rotate, each disc in the FKD 
induces a circumferential shear flow~(Fig.~\ref{fig:fkd_flow}(a)).
The distribution of the shear flow is conveniently 
identified with the strain-rate state of \(\beta\approx 0\) 
(Figs.~\ref{fig:fkd_flow}-\ref{fig:fb_flow}).
Since the circumferential shear flow itself is almost 
unidirectional so that the flow reorientation and line folding 
rarely works.
Then, at the front of the tips, some portion of the fluid 
elements is pushed out to the neighboring 
disc regions~(Fig.~\ref{fig:fkd_flow}(a)), which flow is observed 
as a bifurcating flow~(\(\beta\approx -1\) in 
Fig.~\ref{fig:fkd_flow}(b)).
In contrast, at the back of the tips, the inter-disc
leakage flow from neighboring disc region and circumferential 
flow within the disc converge, which is observed as 
\(\beta\approx 1\) in Figs.~\ref{fig:fkd_flow}(b) and (c).
The bifurcating and converging flows cause the flow reorientation 
and line folding and stretching. Thus, the distribution of them 
gives an essential insight into the flow pattern and resulting 
mixing process.
We note that the development of the bifurcating and converging 
flows is observed in the low-strain-rate regions far from the 
surfaces.
This fact demonstrates that the distribution of the bifurcating 
and converging flows should be responsible for the global flow 
pattern driven by the geometry of FKD and its resulting mixing 
ability.
\newcommand{\figfive}{Snapshots of the flow in FKD with a flow 
rate of 10\;cm\(^{3}\)/s and a screw rotation speed of 200\;rpm: 
(a) velocity field in several axial sections in the kneading 
block zone, (b) the distribution of the strain-rate state, \(\beta\), 
at a mid-section between the barrel and screw surfaces, and (c) 
the strain-rate state distribution at an axial section at the 
center of the third disc in FKD.
}
\begin{figure}[htbp]
\raggedright
\hspace*{5ex}
(a)\\
 \centering
\hspace*{-7ex}
 \includegraphics[width=.65\hsize]{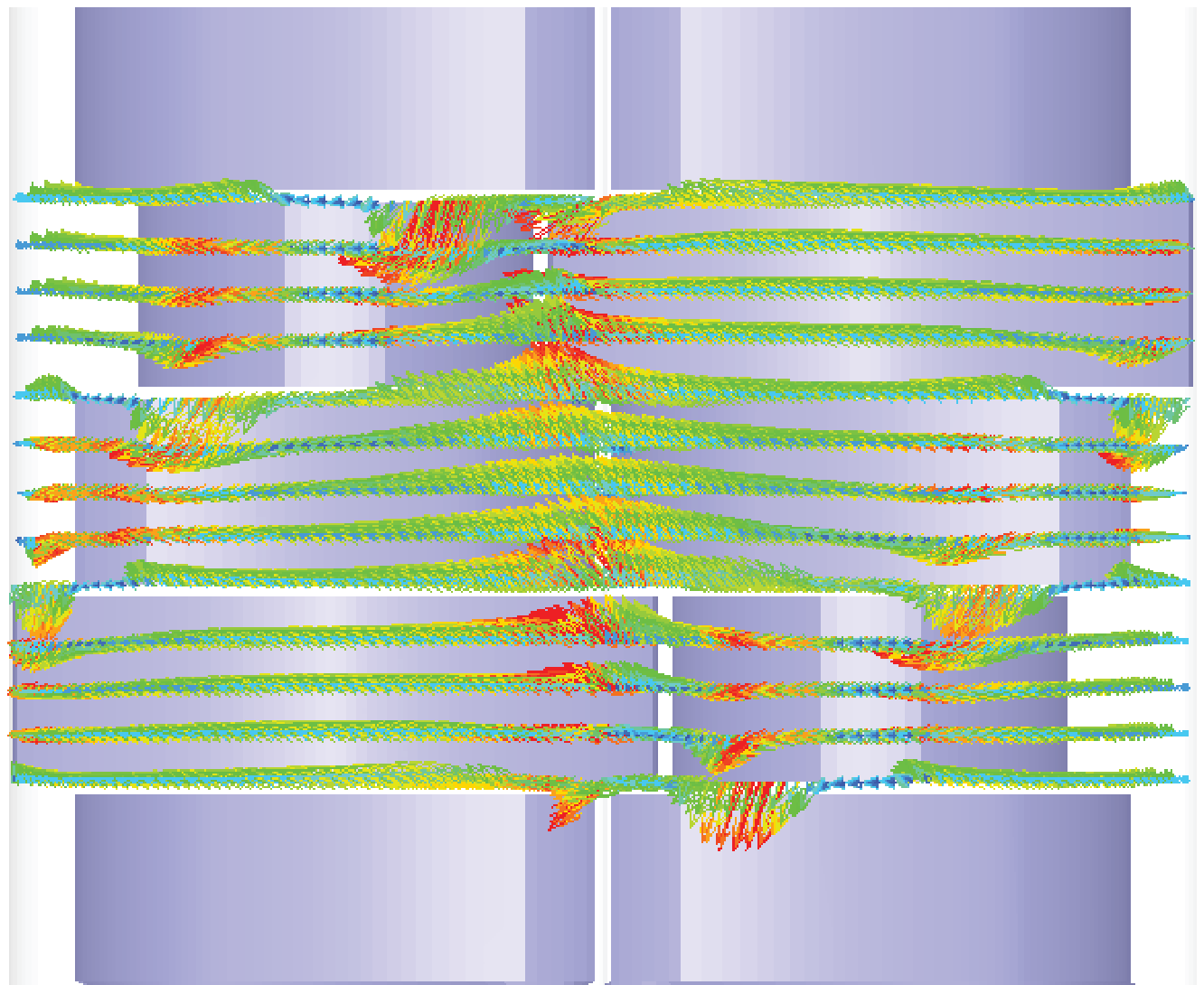}
\\
\raggedright
\hspace*{5ex}
(b)\\
 \centering
 \includegraphics[width=.75\hsize]{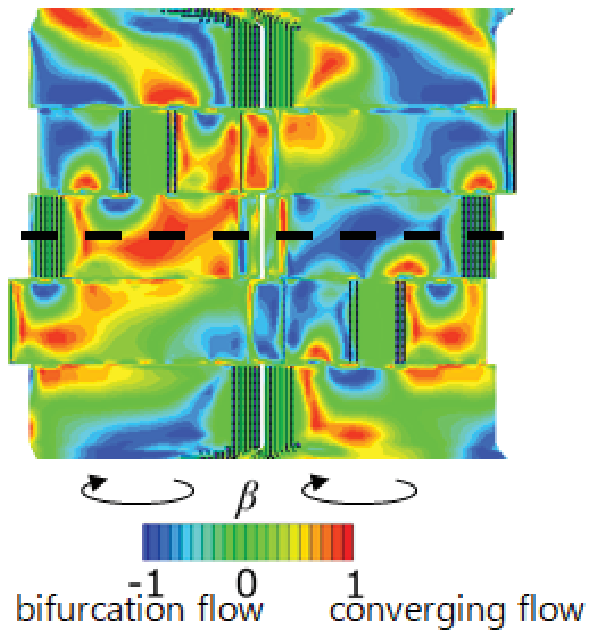}
\\
\raggedright
\hspace*{5ex}
(c)\\
 \centering
 \includegraphics[width=.75\hsize]{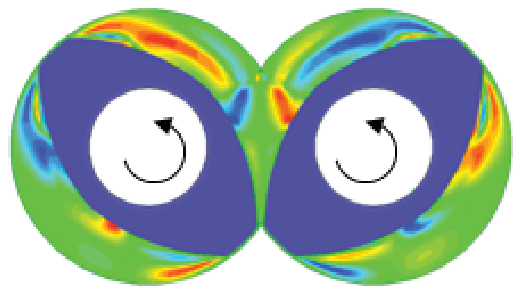}
\caption{
\label{fig:fkd_flow}
\texttt{\figfive}
}
\end{figure}

Next, we discuss how the pitched-tips modify the flow pattern in 
FKD.
For Fs-Ft ptKD, the velocity field and the distribution of the 
strain-rate state are shown in Fig.~\ref{fig:ff_flow}.
We observe in Fig.~\ref{fig:ff_flow}(a) that the bifurcating flow 
through the inter-disc openings is suppressed.
In Figs.~\ref{fig:ff_flow}(b) and (c),  
the volume of the bifurcating and converging flows is reduced 
compared to FKD.
Furthermore, the distribution of \(\beta\) within individual 
disc regions is 
more asymmetric to the extrusion direction than that for FKD: 
within each disc region, the converging flow is localized in the 
downstream region while the bifurcating flow develops in the 
upstream region.
At the same time, the volume of planar shear flow spreads over 
each disc.
These results suggest that in Fs-Ft ptKD, the circumferential 
shear flow is prevailing, leading to less frequent flow 
reorientation and line stretching and folding. 
\newcommand{\figsix}{
Snapshots of the flow in Fs-Ft ptKD with a flow 
rate of 10\;cm\(^{3}\)/s and a screw rotation speed of 200\;rpm: 
(a) velocity field in several axial sections in the kneading block
zone, (b) the distribution of the strain-rate state, \(\beta\), 
at a mid-section between the barrel and screw surfaces, and (c) 
the strain-rate state distribution at an axial section at the 
center of the third disc in Fs-Ft ptKD.
}
\begin{figure}[htbp]
\raggedright
\hspace*{5ex}
(a)\\
 \centering
\hspace*{-7ex}
 \includegraphics[width=.65\hsize]{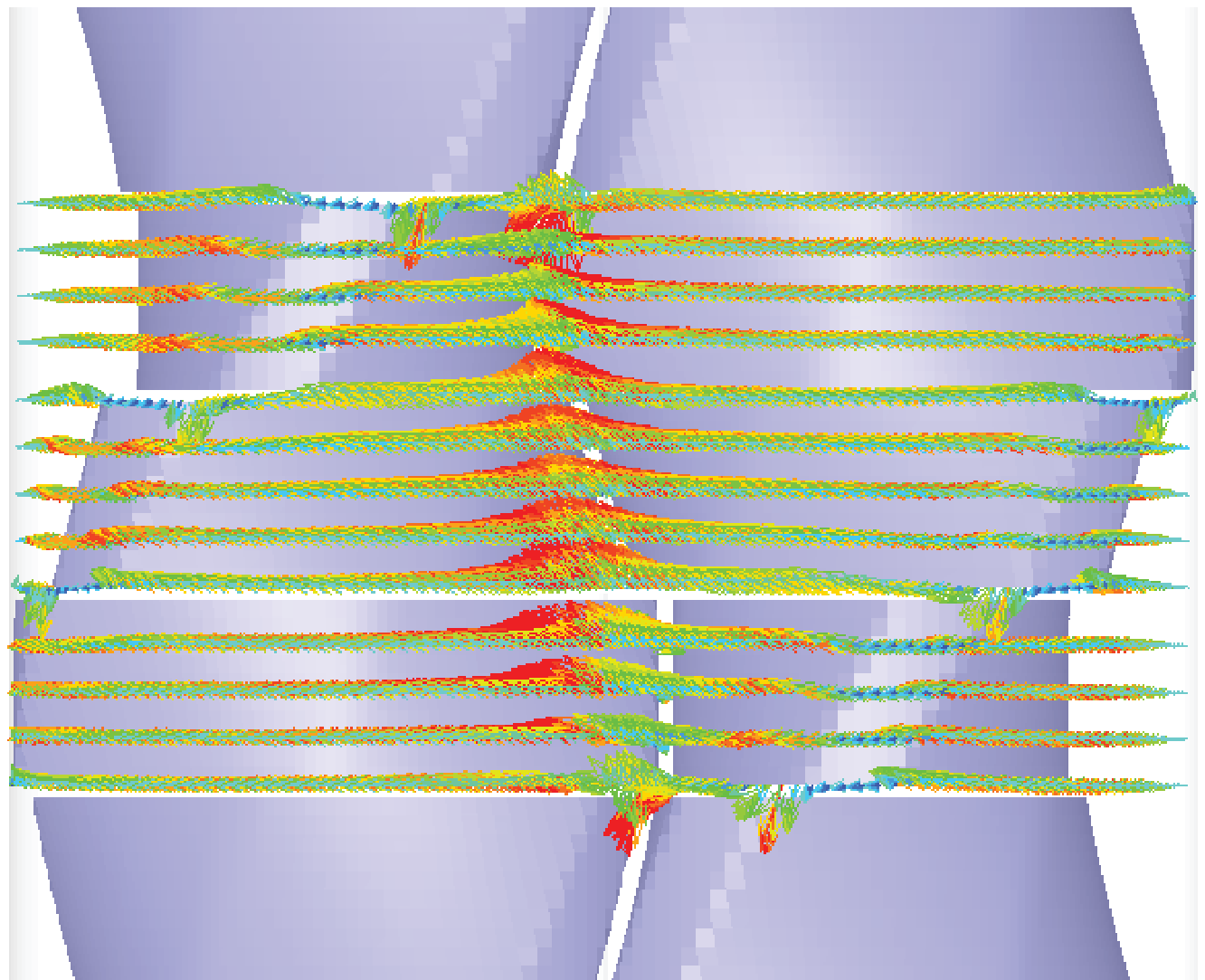}
\\
\raggedright
\hspace*{5ex}
(b)\\
 \centering
 \includegraphics[width=.75\hsize]{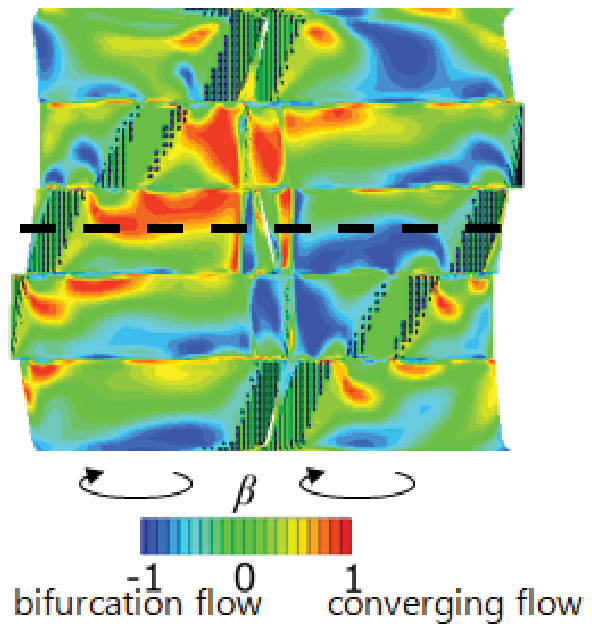}
\\
\raggedright
\hspace*{5ex}
(c)\\
 \centering
 \includegraphics[width=.75\hsize]{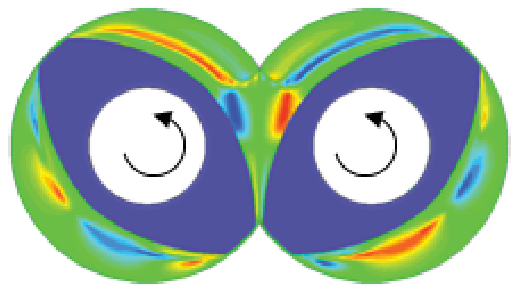}
\caption{
\label{fig:ff_flow}
\texttt{\figsix}
}
\end{figure}

For Fs-Bt ptKD, the velocity field and the distribution of the 
strain-rate state are shown in Fig.~\ref{fig:fb_flow}.
In Fig.~\ref{fig:fb_flow}(a), inter-disc leakage flow as 
well as the circumferential flow in each disc are observed 
in a similar manner to FKD.
Correspondingly, the distribution of the strain-rate state 
is similar to that for FKD.
However, Figs.~\ref{fig:fb_flow}(b) and (c) clearly show that the 
volume of the non-planar flows becomes larger than that in FKD, 
indicating more frequent reorientation when fluid elements pass 
through the kneading block zone of Fs-Bt ptKD compared to FKD.
\newcommand{\figseven}{
Snapshots of the flow in Fs-Bt ptKD with a flow 
rate of 10\;cm\(^{3}\)/s and a screw rotation speed of 200\;rpm: 
(a) velocity field in several axial sections in the kneading block
zone, (b) the distribution of the strain-rate state, \(\beta\), 
at a mid-section between the barrel and screw surfaces, and (c) 
the strain-rate state distribution at an axial section at the 
center of the third disc in Fs-Bt ptKD.
}
\begin{figure}[htbp]
\raggedright
\hspace*{5ex}
(a)\\
 \centering
\hspace*{-7ex}
 \includegraphics[width=.65\hsize]{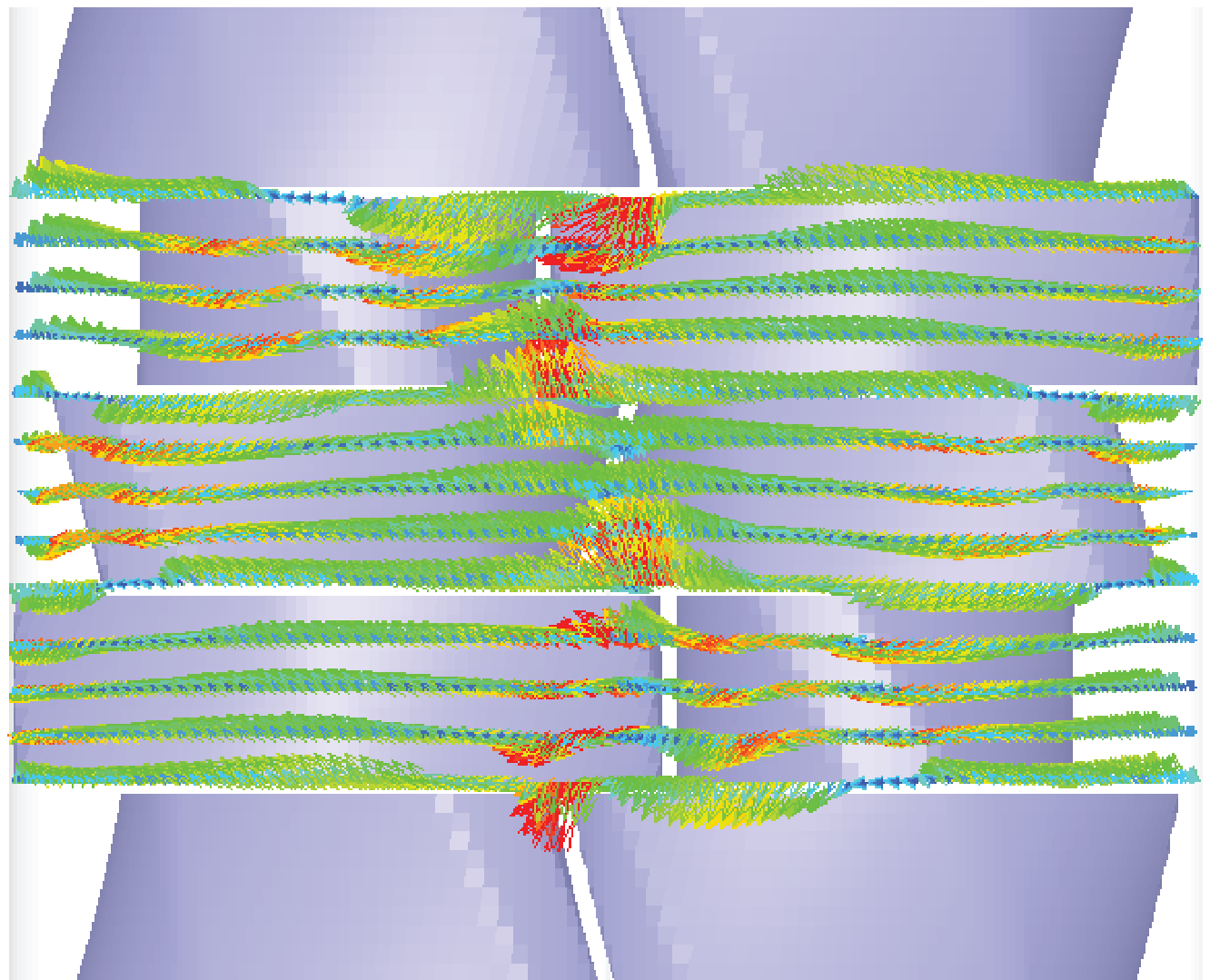}
\\
\raggedright
\hspace*{5ex}
(b)\\
 \centering
 \includegraphics[width=.75\hsize]{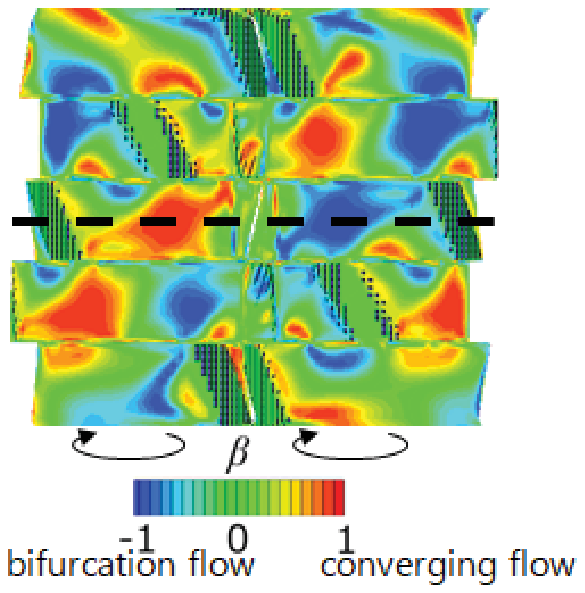}
\\
\raggedright
\hspace*{5ex}
(c)\\
 \centering
 \includegraphics[width=.75\hsize]{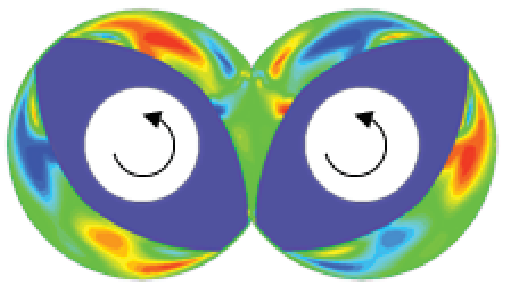}
\caption{
\label{fig:fb_flow}
\texttt{\figseven}
}
\end{figure}

For a more quantitative evaluation of the non-planar flows, we 
compute the section-average of the strain-rate state as a 
function of the axial position.
To separately evaluate the bifurcating and 
converging flows, we decompose \(\beta\) into the following 
quantities,
\begin{align}
 \beta_{+}
&=
\begin{cases}
 \beta & (\beta > 0),
\\
0 & \text{otherwise},
\end{cases}
~~~~
 \beta_{-}
&=
\begin{cases}
 \beta & (\beta < 0),
\\
0 & \text{otherwise},
\end{cases}
\end{align}
where \(\beta_{\pm}\) respectively represents the converging and 
bifurcating flows,
and \(\beta_{+}+\beta_{-}=\beta\) holds.
In Fig.~\ref{fig:axial_beta}, the axial profile of 
\(\langle\beta_{\pm}\rangle_{z}\) is drawn.
Since the flow pattern specific to kneading blocks develops 
at the inside discs out of the five discs shown in 
Fig.~\ref{fig:geom} that are sandwiched between two other discs,
we focus on the variation of \(\langle\beta_{\pm}\rangle_{z}\) 
within the inner three discs region.

For FKD, the converging flow increases at the upstream side 
within individual disc regions while the bifurcating flow 
increases at the downstream side.
This reflects the asymmetry of \(\beta_{\pm}\) with respect to 
the extrusion direction within a disc observed in 
Fig.~\ref{fig:fkd_flow}(b).
For Fs-Ft ptKD, both the converging and bifurcating flows 
are substantially suppressed compared to FKD.
This corresponds to the fact that in Fs-Ft ptKD the planar 
flow is prevailing and the occurrence of the non-planar flow is 
relatively suppressed, as observed in Figs.~\ref{fig:ff_flow}(b) 
and (c),
indicating the bifurcating and converging flow from 
the circumferential flow occurs in limited regions.
In contrast, for Fs-Bt ptKD, both the converging and 
bifurcating flows are enhanced compared to FKD, and their 
distributions are more extended over each disc, reflecting 
the observation of the enhanced occurrence of the non-planar 
flows in Figs.~\ref{fig:fb_flow}(b) and (c), 
indicating frequent occurrence of the flow reorientation by 
bifurcating and converging flow.
This fact suggests that the flow pattern driven by Fs-Bt ptKD is 
more effective in distributive mixing than in FKD.

These results show that the geometrical modification by the 
pitched tips of the conventional FKD significantly affect the flow 
pattern in the low-strain-rate regions, which is effectively 
characterized by the distribution of the strain-rate state.
The distribution of the converging and bifurcating flows is 
strongly associated with the channel geometry.
We further discuss the effects of the variation of the flow 
pattern on the mixing and dispersive ability in the following sections.
\newcommand{\figeight}{Average strain-rate state over a section 
and one screw rotation as a function of  the axial 
position with a flow rate of 10\;cm\(^{3}\)/s and a screw 
rotation speed of 200\;rpm for FKD (black solid line), Fs-Ft 
ptKD (blue dashed line), and Fs-Bt ptKD (red dotted 
line). 
\(\beta_{\pm}\) respectively  represent positive and negative 
values of \(\beta\), which correspond to the converging and 
bifurcating flows.
The vertical dashed lines indicate the locations of the ends of  
the discs in the KDs.}
\begin{figure}[htbp]
\raggedright
 \centering
 \includegraphics[width=1.1\hsize]{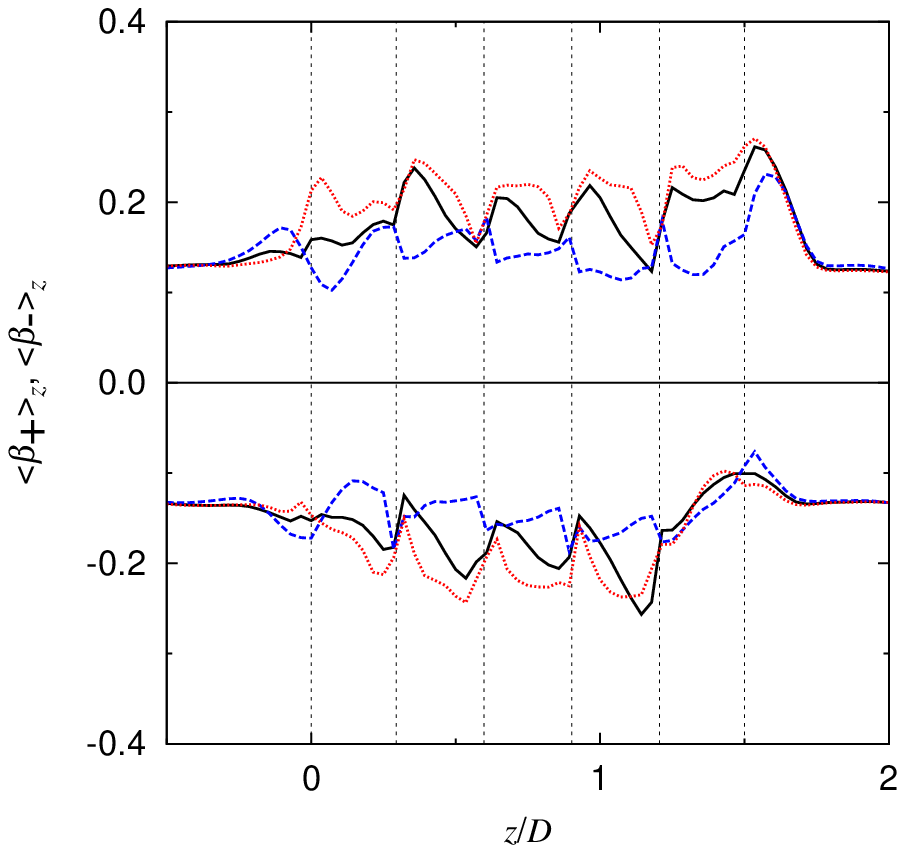}
\caption{
\label{fig:axial_beta}
\texttt{\figeight}
}
\end{figure}

\subsection{Residence time and mean stress during residence}
The distributions of the local residence time and of the 
mean stress during residence in KD zone are 
commonly used to discuss the general characteristics of mixing 
and dispersive abilities in the kneading block zone.
The distribution of the local residence time in the kneading 
block zone reflects the axial mixing ability of the
kneading block, while the distribution 
of the mean stress during residence partly reflects the 
dispersive ability of the kneading block.
The mean stress during the residence of the \(\alpha\)th tracer, 
\(\overline{\sigma_{\alpha}}^{t_{\alpha}}\), is calculated by 
Eq.~(\ref{eq:lagrangian_average_of_f}) by inserting 
\(f=\sigma=\sqrt{(3/2)\tensor{\tau}:\tensor{\tau}}\), which is a 
second-order invariant of the shear stress and measures the 
magnitude of the stress.
Basically, the residence time and 
Eq.~(\ref{eq:lagrangian_average_of_f}) for any \(f\) are defined 
for individual trajectories, and naturally they are correlated.
Accordingly, simultaneous observation of the residence time and 
Eq.~(\ref{eq:lagrangian_average_of_f}) rather than the separate 
observation of them can be more useful to understand the 
properties of the individual trajectories, the ensemble of which 
constitutes the overall mixing process.
For this purpose, we consider the statistical distribution of the 
two quantities,  
\(P(t_{\alpha},\overline{f_{\alpha}}^{t_{\alpha}})\), which is 
called a joint probability density function (joint PDF).
The joint PDF reflects the correlation between 
the two quantities considered, and consequently characterizes the 
overall mixing process.
A separate distribution like the residence time distribution, 
which has been often discussed, is obtained by integrating the 
other variable of the joint PDF as 
\(P(t_{\alpha})=\int_{-\infty}^{\infty}\upd y P(t_{\alpha},y)\).

Figure~\ref{fig:rt_sigma} shows the joint PDF of the residence 
time and \(\overline{\sigma_{\alpha}}^{t_{\alpha}}\) in the 
kneading block zone under a flow 
rate of 10\;cm\(^{3}\)/s and a screw rotation speed of 200\;rpm.
In the case of the FKD shown in Fig.~\ref{fig:rt_sigma}(a), when 
only looking at the residence time distribution, the normalized 
residence time has a wide distribution between 0.3 to 1.6, 
indicating a high capability of axial mixing.
However, the joint PDF of \(t_{\alpha}\) and 
\(\overline{\sigma_{\alpha}}^{t_{\alpha}}\) for FKD has 
two peaks, suggesting that the trajectories are classified 
into two different groups which are not mutually mixed.
One group of the shorter residence time takes smaller 
\(\overline{\sigma_{\alpha}}^{t_{\alpha}}\)
 but 
another group of the longer residence time takes larger 
\(\overline{\sigma_{\alpha}}^{t_{\alpha}}\).
This fact indicates that although the residence time fluctuation 
in FKD is broad, the overall mixing by FKD can have a large 
inhomogeneity.

In contrast, for Fs-Ft and Fs-Bt ptKDs, the joint PDFs of the 
residence time and \(\overline{\sigma_{\alpha}}^{t_{\alpha}}\) 
are unimodal (Figs.~\ref{fig:rt_sigma}(b) and (c)), suggesting 
that the segregation of the trajectories observed for FKD is 
suppressed. 
Quantitatively, the fluctuation of the residence time and the 
level of \(\overline{\sigma_{\alpha}}^{t_{\alpha}}\) are 
different for Fs-Ft and Fs-Bt ptKDs.
For Fs-Ft ptKD, while the mean stress during residence remains almost 
at the same level as for FKD, the normalized residence time lies 
between 0.7 to 1.3, so that the fluctuation of the residence time 
is reduced.
For Fs-Bt ptKD, with the same level of residence time 
fluctuation as observed for FKD, 
a higher level of the mean stress during residence is achieved.
These facts demonstrate that irrespective of the tip angle, the 
pitched tips on FKD are effective at reducing the inhomogeneity of 
the trajectories, and that the mixing characteristics can be 
tunable by the pitched-tip angle as well.
\newcommand{\fignine}{Joint probability density of the residence 
time and the mean stress during residence in the kneading block zone 
with a flow rate of 10\;cm\(^{3}\)/s and a screw rotation speed 
of 200\;rpm: (a) FKD, (b) Fs-Ft ptKD, and (c) Fs-Bt ptKD.
The contour lines drawn in panels (b) and (c) are of the 
probability density of FKD  shown in panel(a).
}
\begin{figure}[htbp]
\raggedright
 \centering
 \includegraphics[width=1.\hsize]{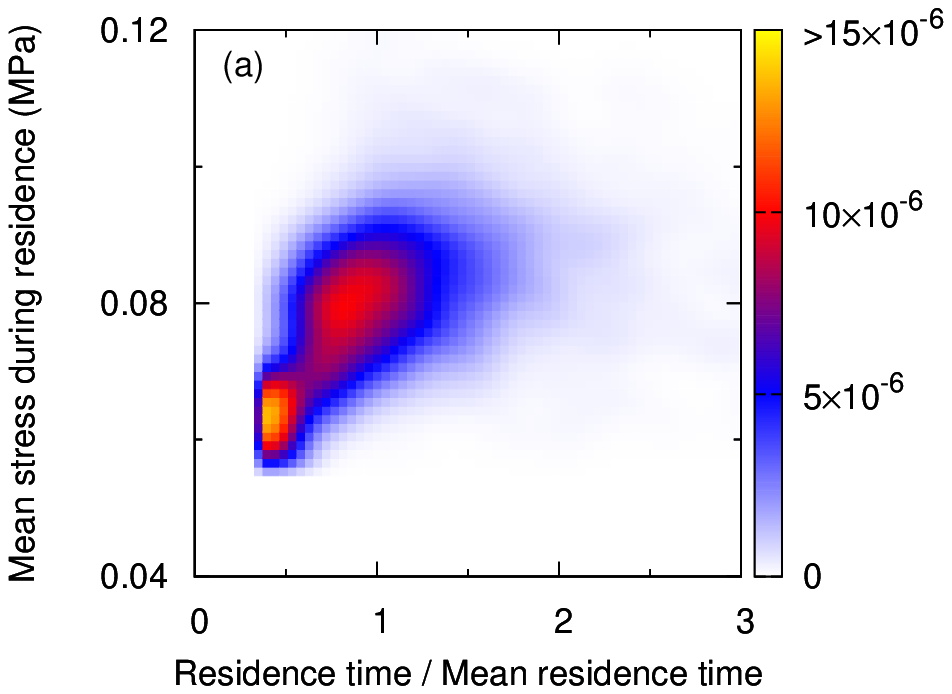}
 \includegraphics[width=1.\hsize]{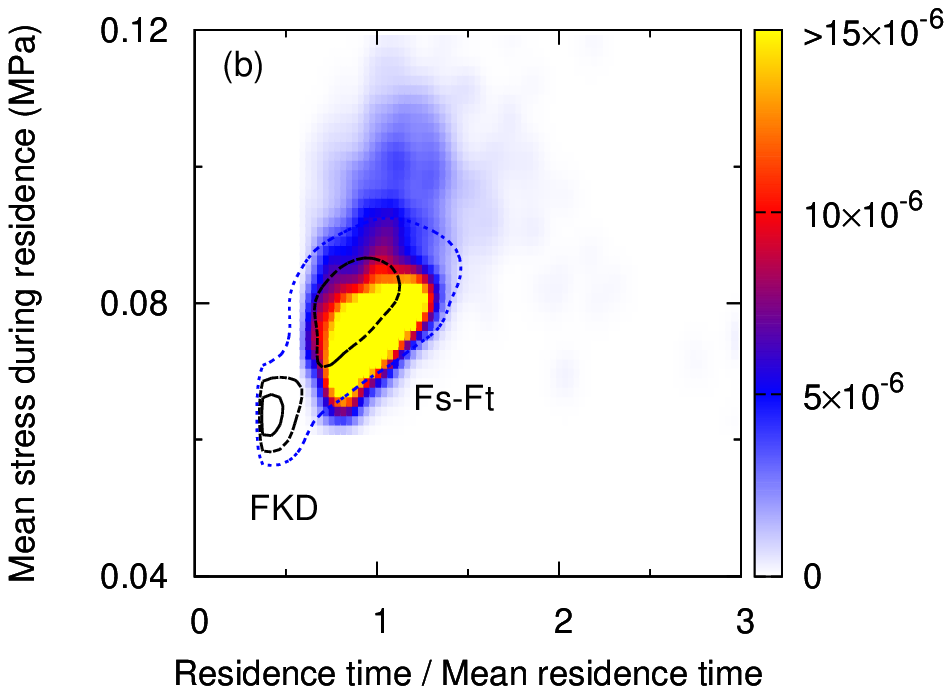}
 \includegraphics[width=1.\hsize]{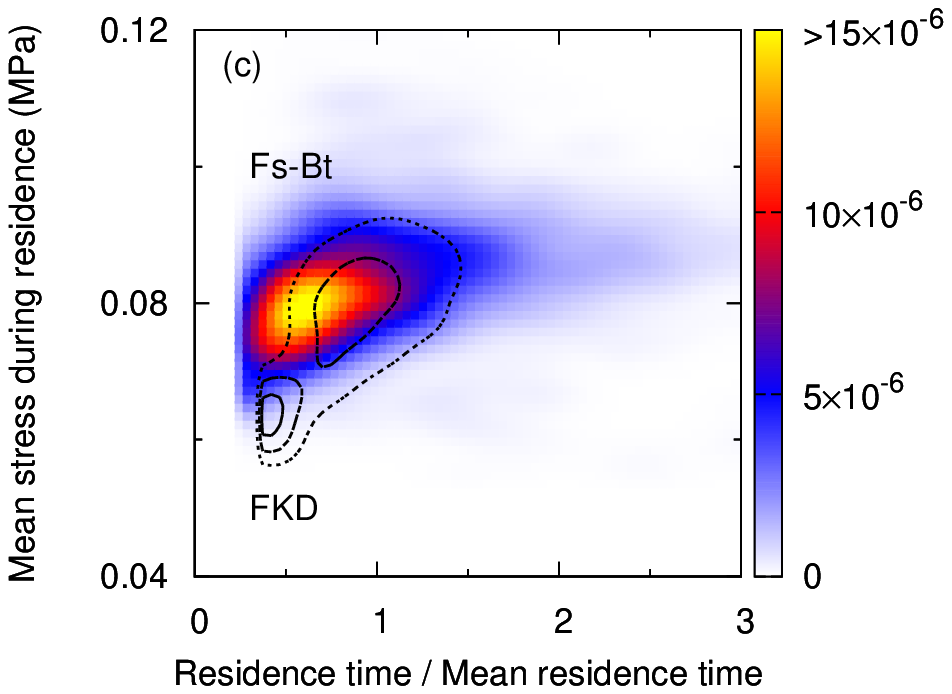}
\caption{
\label{fig:rt_sigma}
\texttt{\fignine}
}
\end{figure}

\subsection{Quantification of mixing: Finite-time Lyapunov exponent}
Although the residence time distribution has been often used to 
assess the mixing ability, it only measures the residence time of 
an individual trajectory, so that it is not a direct measure of 
the mixing process.
In order for mixing to occur, the disappearance of positional 
correlation with time is a necessary process.
For the nonrecurrent dynamics occurring in extrusion, this 
process can be quantified by the finite-time Lyapunov 
exponent~(FTLE).
The FTLE measures the exponential growth rate over a finite time 
interval of the distance between two initially nearby points, and 
is defined as 
\begin{align}
\label{eq:ftle}
 \lambda_{t}(t_{0}) &= \frac{1}{t}\ln\frac{\left|\vec{l}(t_{0}+t)\right|}{\left|\vec{l}(t_{0})\right|},
\end{align}
where \(\vec{l}\) is the relative position vector between the two 
points, \(t_{0}\) is the initial time, and \(t\) is the time 
interval.
When \(\lambda_{t}\) is positive, 
the disappearance of positional correlation is faster for larger 
\(\lambda_{t}\), implying a high mixing ability.
But zero or negative \(\lambda_{t}\) indicates the maintenance of 
the positional correlation, implying a poor mixing ability.
It is supposed that 
The line stretching by elongational flow 
and reorientation flows are supposed to cause a positive FTLE, 
whereas unidirectional planar flow is less likely to cause a 
large FTLE.

Figure~\ref{fig:ftle} shows the PDF of the time-averaged FTLE during 
residence in the kneading block zone, 
\(\overline{\lambda_{t,\alpha}}^{t_{\alpha}}\), under a flow rate 
of 10\;cm\(^{3}\)/s and a screw rotation speed of 200\;rpm.
The time-averaged FTLE during residence is calculated for each 
trajectory specified by \(\alpha\) with 
Eq.~(\ref{eq:lagrangian_average_of_f}) by inserting 
Eq.~(\ref{eq:ftle}) into \(f\). The individual 
\(\overline{\lambda_{t,\alpha}}^{t_{\alpha}}\) characterizes the 
mixing ability along the \(\alpha\)th trajectory.
In order to discuss the overall mixing ability of the kneading 
block zone, 
PDF of \(\overline{\lambda_{t,\alpha}}^{t_{\alpha}}\) is 
calculated.
For FKD, we observe a bimodal distribution of 
\(\overline{\lambda_{t,\alpha}}^{t_{\alpha}}\), which is composed 
of a larger fraction of positive FTLE and a smaller fraction 
of negative FTLE.
The large fraction of the trajectories with positive FTLE 
basically indicates that there is some level of good mixing ability of FKD. 
However, at the same time, the finite fraction of negative FTLE 
clearly indicates the existence of some portion with poor mixing 
ability, which should lead to a large inhomogeneity of the 
mixing process, and can be a bottleneck in the overall mixing 
process.
\newcommand{\figten}{Probability density of the mean FTLE 
during residence with a  flow rate of 10\;cm\(^{3}\)/s and a 
screw rotation speed of 200\;rpm.}
\begin{figure}[htbp]
\raggedright
 \centering
 \includegraphics[width=1.1\hsize]{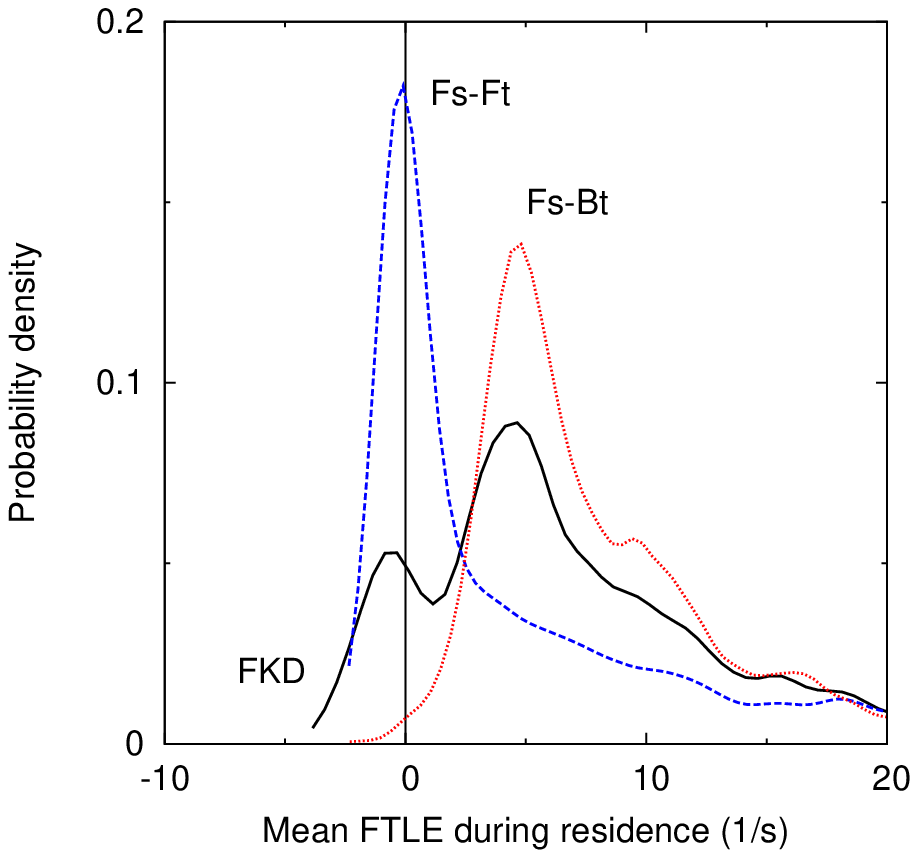}
\caption{
\label{fig:ftle}
\texttt{\figten}
}
\end{figure}

Figures~\ref{fig:fkd_low_ftle} and \ref{fig:fkd_high_ftle} show 
typical pathlines of two tracers of initially nearby locations: 
Fig.~\ref{fig:fkd_low_ftle} is for a case with a negative FTLE 
and Fig.~\ref{fig:fkd_high_ftle} is for a case with a positive 
FTLE.
In the case with a negative FTLE, the initially nearby two 
tracers remain close to each other while passing through the 
kneading block zone.
In contrast, for the case with a positive FTLE, the two initially 
nearby tracers start to move apart at a certain point, and 
after that they follow completely uncorrelated trajectories.

The trajectories with large positive FTLE are found to move back 
and forth over discs more frequently than those with negative 
FTLE~(Fig.~\ref{fig:fkd_high_ftle}).
The back and forth motion between discs is associated with the 
bifurcating and converging flows observed in Fig.\ref{fig:fkd_flow}.

For Fs-Ft and Fs-Bt ptKDs, although the levels of the FTLE are 
different, the distributions of 
\(\overline{\lambda_{t,\alpha}}^{t_{\alpha}}\) are almost 
unimodal.
For Fs-Ft ptKD, although the average value of 
\(\overline{\lambda_{t,\alpha}}^{t_{\alpha}}\)
over the ensemble of the trajectories
is positive, a 
large fraction of the trajectories takes a value around zero, 
indicating that the mixing ability is rather suppressed compared 
to FKD.
For Fs-Bt ptKD, the fraction of negative values of 
\(\overline{\lambda_{t,\alpha}}^{t_{\alpha}}\) almost 
vanishes. 
The average 
\(\overline{\lambda_{t,\alpha}}^{t_{\alpha}}\) 
over the trajectory ensemble
takes a larger 
value than that for FKD, suggesting an enhanced mixing ability of 
Fs-Bt ptKD.

\newcommand{\figeleven}{(a) Typical pathlines of a pair of 
tracers at initially nearby locations at the inlet of FKD where 
one tracer is chosen to take a negative mean FTLE during 
residence.
The horizontal dashed lines indicate the locations of the ends of 
the discs of FKD.
(b) The initial location of the two trajectories in (a) at the 
inlet section of the kneading block zone. The distance between the two 
particles is 0.0062\(D\).
}
\begin{figure}[htbp]
\raggedright
 \centering
 \includegraphics[width=1.\hsize]{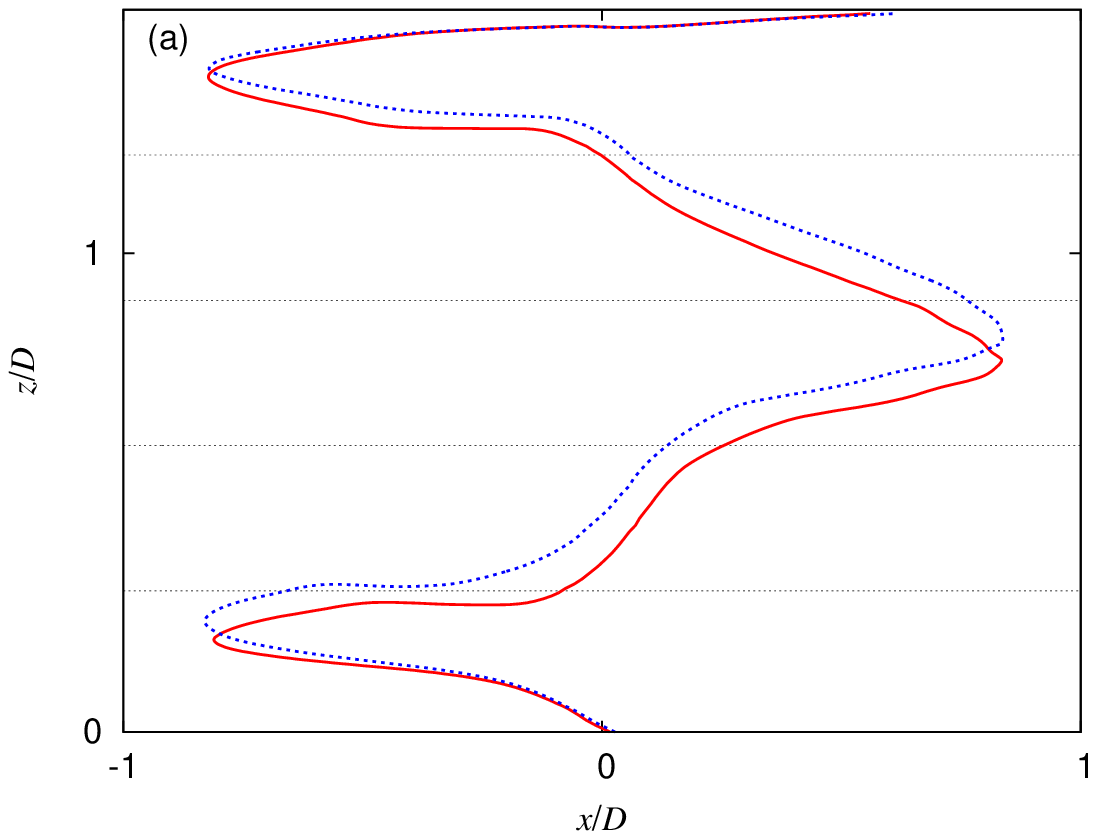}
 \includegraphics[width=1.\hsize]{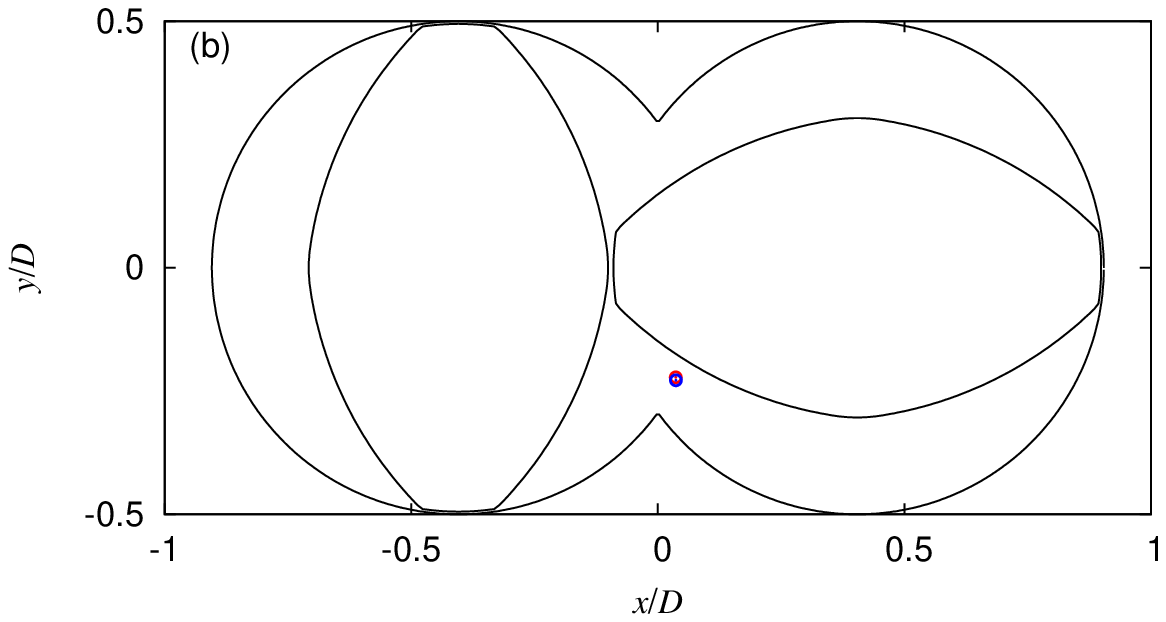}
\caption{
\label{fig:fkd_low_ftle}
\texttt{\figeleven}
}
\end{figure}

\newcommand{\figtwelve}{(a) Typical pathlines of a pair of 
tracers at initially nearby locations at the inlet of FKD where 
one tracer is chosen to take a positive mean FTLE during 
residence.
The horizontal dashed lines indicate the locations of the ends of 
the discs of FKD.
(b) The initial location of the two trajectories in (a) at the 
inlet section of the kneading block zone. The distance between the two 
particles is 0.0075\(D\).
}
\begin{figure}[htbp]
\raggedright
 \centering
 \includegraphics[width=1.\hsize]{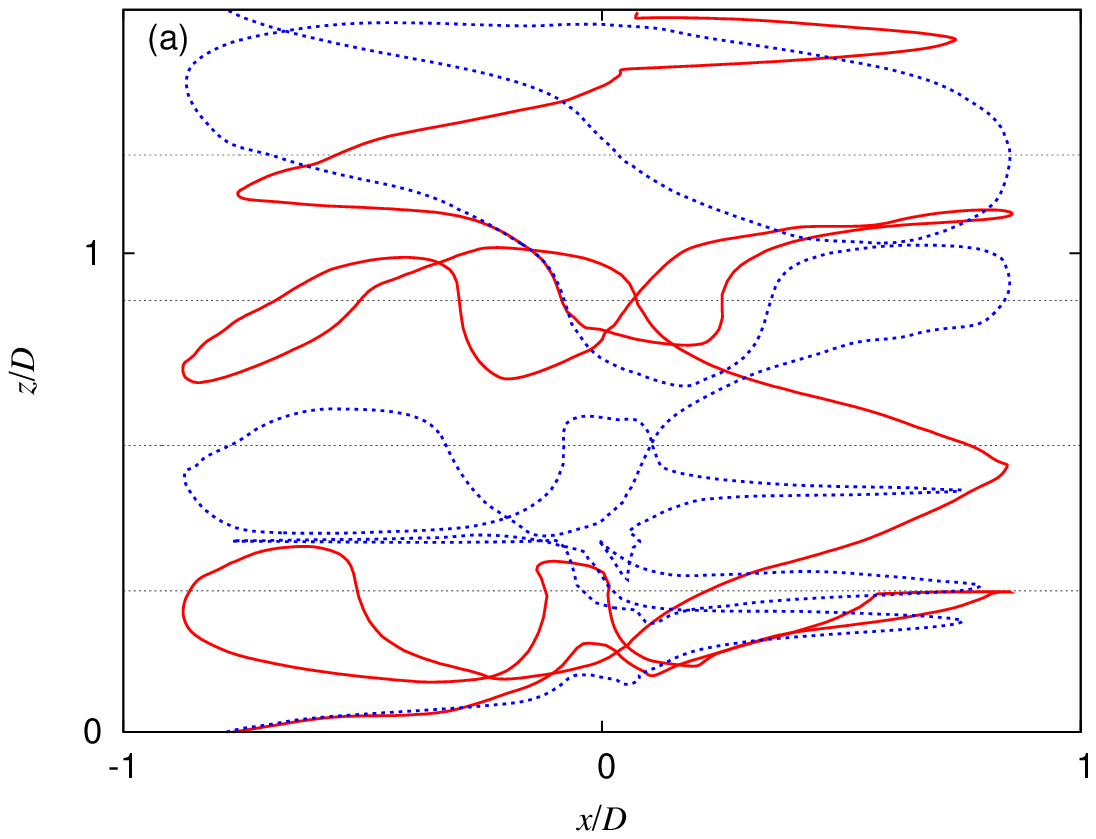}
 \includegraphics[width=1.\hsize]{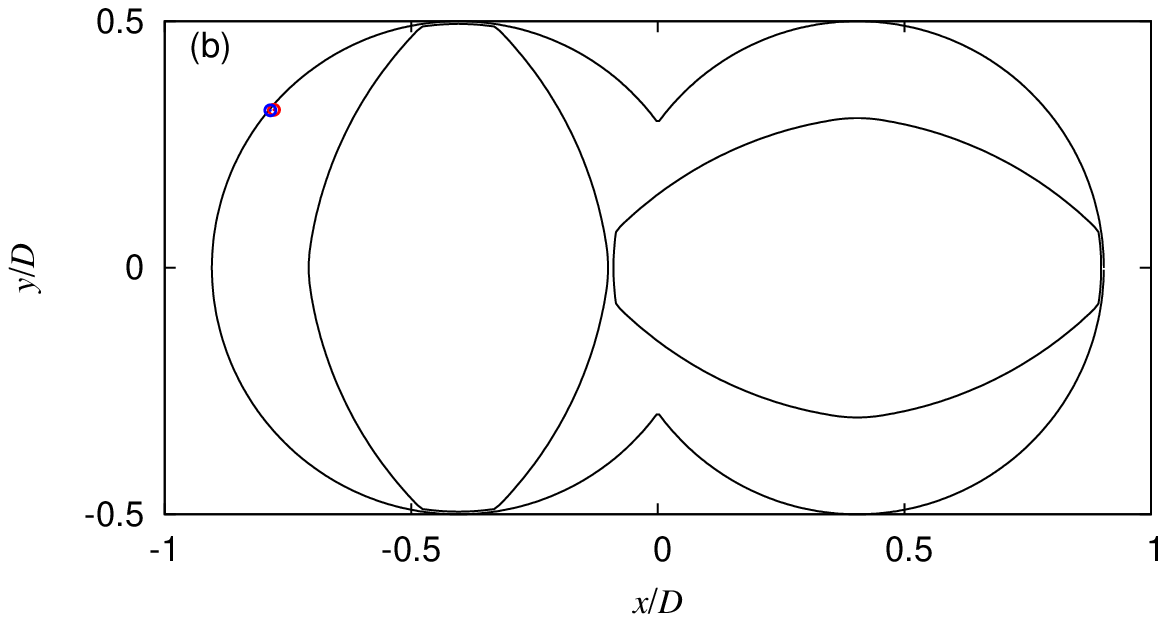}
\caption{
\label{fig:fkd_high_ftle}
\texttt{\figtwelve}
}
\end{figure}

Figure~\ref{fig:ftle_sigma} shows the joint PDF of 
\(\overline{\lambda_{t,\alpha}}^{t_{\alpha}}\) and 
\(\overline{\sigma_{\alpha}}^{t_{\alpha}}\) under a flow rate of 
10\;cm\(^{3}\)/s and a screw rotation speed of 200\;rpm.
Figure~\ref{fig:ftle_sigma}(a), for FKD, clearly shows that the 
trajectories with lower mean stress take negative or positive but 
small values of FTLE, suggesting that the poor mixing ability of 
this group causes a large inhomogeneity in the overall mixing by FKD.
For Fs-Ft ptKD (Fig.~\ref{fig:ftle_sigma}(b)) and Fs-Bt 
ptKD (Fig.~\ref{fig:ftle_sigma}(c)), the distributions of 
\(\overline{\lambda_{t,\alpha}}^{t_{\alpha}}\) and 
\(\overline{\sigma_{\alpha}}^{t_{\alpha}}\) are unimodal, so that 
the mixing processes in these ptKDs are qualitatively rather 
homogeneous.
These results reveal the suppressed mixing in Fs-Ft ptKD 
and the enhanced mixing in Fs-Bt ptKD compared to FKD.
The distribution of the FTLE turns out to be useful to quantify the 
mixing characteristics of different mixing elements.
\newcommand{\figthirteen}{Joint probability density of 
the mean FTLE during residence and the mean stress during 
residence in the kneading block zone with a flow rate of 
10\;cm\(^{3}\)/s and a screw rotation speed of 200\;rpm: (a) FKD, 
(b) Fs-Ft ptKD, and (c) Fs-Bt ptKD.}
\begin{figure}[htbp]
\raggedright
 \centering
 \includegraphics[width=.97\hsize]{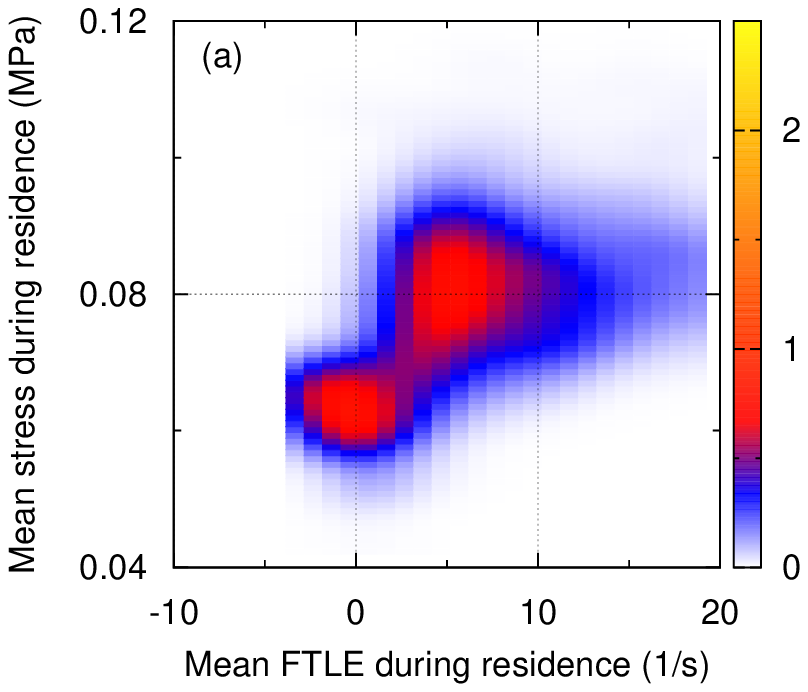}
 \includegraphics[width=.97\hsize]{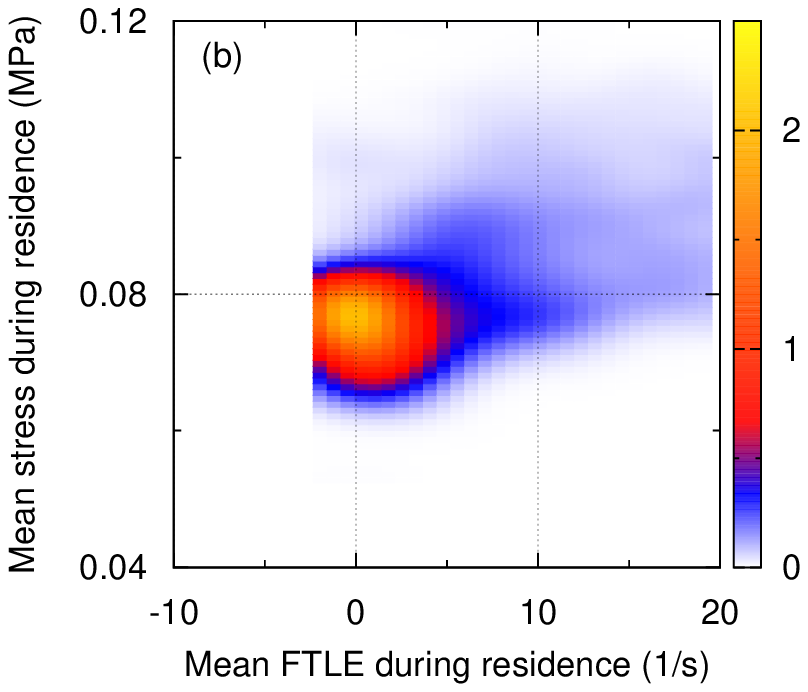}
 \includegraphics[width=.97\hsize]{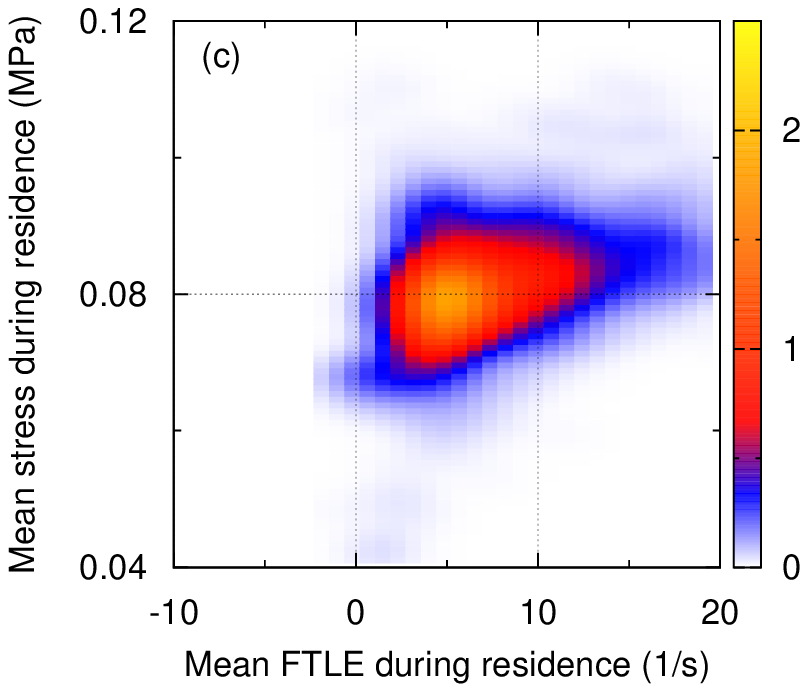}
\caption{
\label{fig:ftle_sigma}
\texttt{\figthirteen}
}
\end{figure}

\subsection{Quantification of dispersive mixing efficiency}
Dispersive mixing is mainly promoted when a fluid element 
passes through the high-stress regions.
In twin-screw mixers, the highest level of the stress is achieved 
in the small-gap regions, including the tip--barrel clearance and 
the inter-meshing region, independent of the disc-stagger angle 
and the pitched-tip angle.
Figure~\ref{fig:high_stress_region}(a) shows a typical time 
series of the stress magnitude on a tracer.
The impulsive peaks in Fig.~\ref{fig:high_stress_region}(a) 
indicate that the tracer passes through the high-stress region.
Although the passage time of the high-stress regions is highly 
limited in the whole residence time in the kneading block zone, it is a 
determining factor for dispersion efficiency.
Since the passage of the fluid to the high-stress regions 
depends on the flow pattern in the low-stress regions, it should 
be directly affected by the geometry of the mixing elements.
To quantify the dispersion efficiencies of the different 
KDs, we computed the passage time in the high-stress regions.

In order to define the high-stress regions, we introduce a 
threshold value, \(\sigma_{c}\), for the stress magnitude, which 
value was  arbitrarily set to 0.2\;MPa, indicated in 
Fig.~\ref{fig:high_stress_region}(a), to cover both the 
tip--barrel clearance and the inter-meshing region.
Based on this,
the passage time in the high-stress regions for the \(\alpha\)th 
tracer, \(t_{h,\alpha}\), is calculated as
\begin{align}
t_{h,\alpha} &= \int_{0}^{t_{\alpha}}\upd u
\int\upd\vec{x}\delta\left(\vec{x}-\vec{X}_{\alpha}(u)\right)
\theta(\sigma(\vec{x},u)-\sigma_{c}),
\end{align}
where \(\theta(...)\) represents Heaviside step function. The 
\(t_{h,\alpha}\) is the summation of the time when \(\sigma\) 
exceeds the threshold \(\sigma_{c}\) for the \(\alpha\)th tracer.

Figure~\ref{fig:high_stress_region}(b) shows the PDF of 
\(t_{h,\alpha}\) under a flow rate of 10\;cm\(^{3}\)/s and a 
screw rotation speed of 200\;rpm.
For both Fs-Ft and Fs-Bt ptKDs, the fraction of the vanishing 
passage time in the high-stress regions is slightly larger than 
for FKD. 
However, the distribution of the finite time passage of the 
high-stress regions is almost similar for FKD, Fs-Ft and Fs-Bt 
ptKDs, suggesting that the dispersion efficiencies for Fs-Ft and 
Fs-Bt ptKDs are at the same level as for FKD.

\newcommand{\figfourteen}{(a) Typical time series of the stress magnitude experienced by a 
tracer. 
An arbitrarily set threshold value drawn by a horizontal line 
defines the high-stress region, which almost covers the  
tip--barrel clearance regions and the inter-meshing zone.
(b)
Probability density of the relative residence time of the 
high-stress region with a flow rate of 10\;cm\(^{3}\)/s and a 
screw rotation speed of 200\;rpm.
}
\begin{figure}[htbp]
\raggedright
 \centering
 \includegraphics[width=1.1\hsize]{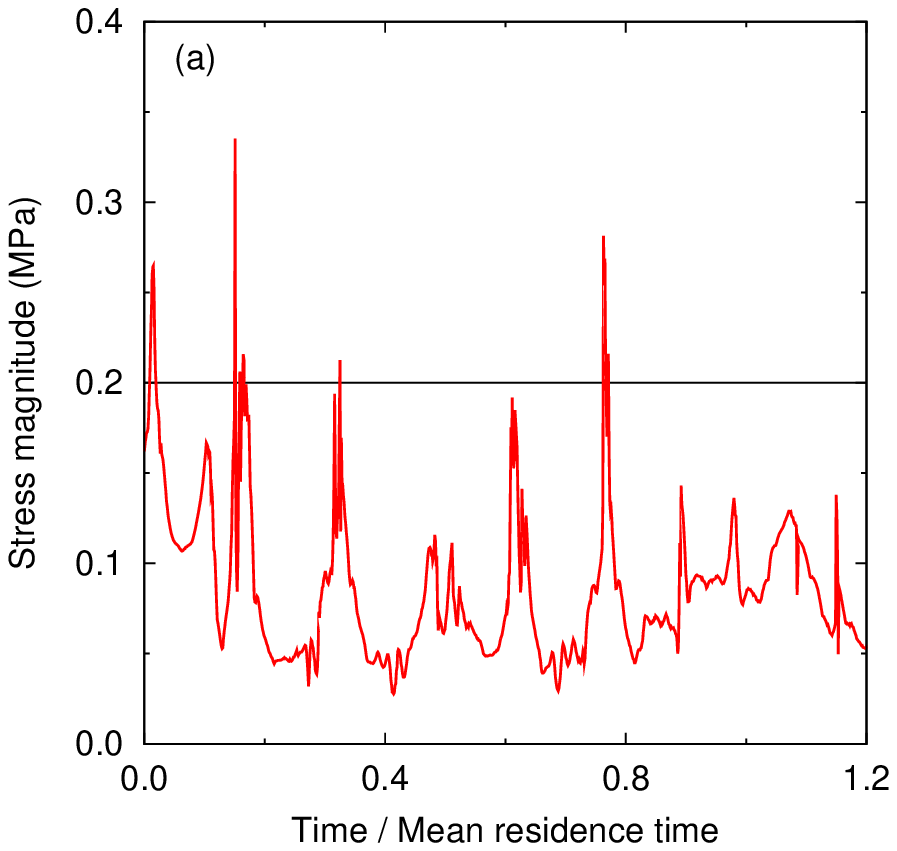}
 \includegraphics[width=1.1\hsize]{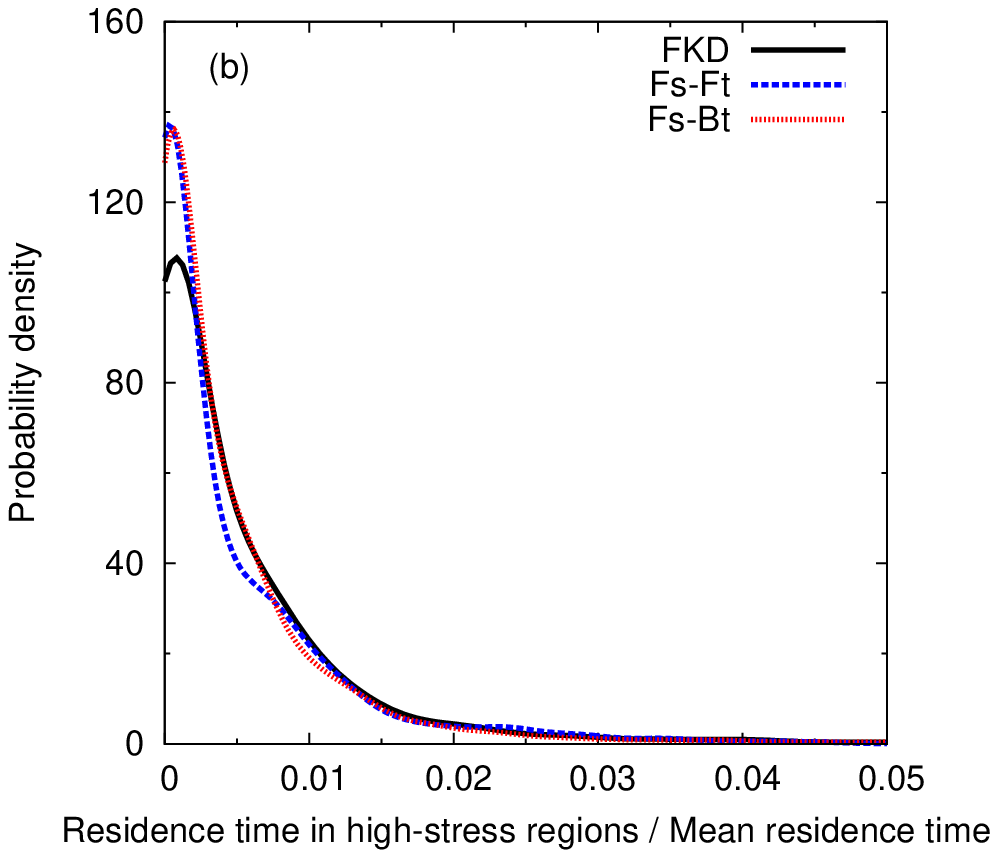}
\caption{
\label{fig:high_stress_region}
\texttt{\figfourteen}
}
\end{figure}

\section{Conclusions}
For mixing elements in twin-screw extrusion, we investigated the 
effects of a geometrical modification by pitched tips of the 
conventional forward kneading block (FKD) on the flow pattern 
and the mixing characteristics, employing numerical simulations of 
the three-dimensional melt flow.
Based on the FKD geometry, two different pitched tips, forward
and backward pitched-tip angles, were considered, and these 
two pitched-tip kneading block~(ptKD) are respectively called 
Fs-Ft and Fs-Bt ptKDs.
In flow driven by FKD, we found that the trajectories were 
composed of two qualitatively different groups: poorly mixing ones 
and highly mixing ones. 
The poorly mixing group causes a large inhomogeneity of the 
overall mixing in FKD.
In contrast, in flows driven by either ptKD, the mixing 
characteristics are found to be qualitatively homogeneous.

The mixing characteristics depend on the geometry of the ptKD.
Fs-Ft ptKD has a suppressed level of mixing compared to FKD, but 
has the same level of dispersion ability as FKD.
Fs-Bt ptKD has a highly enhanced mixing ability as well as 
dispersion ability.
Our results revealed that the geometrical modification by the 
pitched tips of the conventional KD is effective at improving, as 
well as tuning, the mixing characteristics of KD.

In the investigation of the mixing characteristics in this paper, 
we employed different approaches, including the strain-rate 
state distribution, the distribution of the finite-time Lyapunov 
exponent (FTLE), and the passage time in the high-stress region, 
in addition to the conventional approaches like the residence 
time distribution.
These approaches were found to be useful for clarifying the relations 
between the channel geometry, the flow pattern,  and the mixing 
characteristics.
Understanding the relation between the channel geometry and the 
flow pattern behind different mixing characteristics of different 
mixing elements is an essential issue in the optimization and the 
development of mixing processes.
In this direction, our approaches can be applied to other mixing 
devices, including twin-screw ones as well as single screw ones.

\bigskip

\section*{Acknowledgments}
%
The numerical calculations were partly carried out using the
computer facilities at the Research Institute for Information Technology
at Kyushu University.
This work has been supported by Grants-in-Aid for Scientific Research
(JSPS KAKENHI) under Grants Nos.~JP26400433, and JP15H04175.
Financial support from Hosokawa Powder Technology Foundation is 
also greatfully acknowledged.

\appendix
\section{Strain-rate state and local flow pattern}
\label{sec:stran_state}
The local flow pattern is closely related to the 
distribution of the strain-rate state. For instance, the 
biaxial elongational flow is associated with the bifurcation of 
the flow, while the uniaxial elongational (biaxial compression) 
flow is associated with the convergence of the flow.
Since the flow pattern is mainly controlled by the channel 
geometry, the spatial distribution of the strain-rate state is 
useful for understanding the relation between the geometry and 
the structure of the local flow pattern~\cite{Nakayama2016Strain}.
The strain-rate state distribution has also been used to 
discuss the elongational flow distribution in 
turbulence~\cite{Lund1994Improved,Toonder1996Criterion}.

We here briefly describe the quantification of the strain-rate 
state.
The strain-rate state is evaluated with the following 
dimensionless scalar,
\begin{align}
\label{eq:beta}
 \beta 
&=\frac{3\sqrt{6}\det\tensor{D}}
{\left(\tensor{D}:\tensor{D}\right)^{3/2}}.
\end{align}
which take a value in \([-1,1]\).
\(\beta\) is an invariant of \(\tensor{D}\) since both 
\(\det\tensor{D}\) and \(\tensor{D}:\tensor{D}\) are invariants 
of \(\tensor{D}\).
For an arbitrary uniaxial elongational flow, \(\beta\) takes a 
positive value, while for an arbitrary biaxial elongational flow, 
\(\beta\) takes a negative value.
Special cases are \(\beta=1\) for equibiaxial compression 
(uniaxial elongational) flow, \(\beta=-1\) for equibiaxial 
elongational flow, and \(\beta=0\) for planar shear flow.
Therefore, a positive value of \(\beta\) indicates a converging 
flow, while a negative value of \(\beta\) indicates a bifurcating flow.


\bigskip

\renewcommand{\refname} {Literature Cited}


%
\end{document}